\documentclass[%
 reprint,
 amsmath,amssymb,
 aps,
]{revtex4-1}

\usepackage{graphicx}
\usepackage{dcolumn}
\usepackage{bm}
\usepackage{hyperref}
\usepackage{float}
\usepackage{tabularx}
\usepackage{setspace}

\usepackage{xcolor}
\newcommand{\editcolor}[0]{black}

\usepackage{xr}
\makeatletter
\newcommand*{\addFileDependency}[1]{
	\typeout{(#1)}
	\@addtofilelist{#1}
	\IfFileExists{#1}{}{\typeout{No file #1.}}
}
\makeatother
\newcommand*{\myexternaldocument}[1]{%
	\externaldocument{#1}%
	\addFileDependency{#1.tex}%
	\addFileDependency{#1.aux}%
}
\myexternaldocument{SI}

\usepackage{appendix}

\begin{document}
	
	\singlespacing
	
	\title{Pattern formation in odd viscoelastic fluids}
	\author{Carlos Floyd}
	\affiliation{The Chicago Center for Theoretical Chemistry, The University of Chicago, Chicago, Illinois 60637, USA
	}
	\affiliation{Department of Chemistry, The University of Chicago, Chicago, Illinois 60637, USA
	}
	\affiliation{The James Franck Institute, The University of Chicago, Chicago, Illinois 60637, USA
	}
	\author{Aaron R.\ Dinner}
	\affiliation{The Chicago Center for Theoretical Chemistry, The University of Chicago, Chicago, Illinois 60637, USA
	}
	\affiliation{Department of Chemistry, The University of Chicago, Chicago, Illinois 60637, USA
	}
	\affiliation{The James Franck Institute, The University of Chicago, Chicago, Illinois 60637, USA
	}
	\author{Suriyanarayanan Vaikuntanathan}
	\email{svaikunt@uchicago.edu}
	\affiliation{The Chicago Center for Theoretical Chemistry, The University of Chicago, Chicago, Illinois 60637, USA
	}
	\affiliation{Department of Chemistry, The University of Chicago, Chicago, Illinois 60637, USA
	}
	\affiliation{The James Franck Institute, The University of Chicago, Chicago, Illinois 60637, USA
	}

	\date{\today}
	\begin{abstract}
		Non-reciprocal interactions fueled by local energy consumption can be found in biological and synthetic active matter at scales where viscoelastic forces are important. Such systems can be described by ``odd'' viscoelasticity, which assumes fewer material symmetries than traditional theories.  Here we study odd viscoelasticity analytically and using lattice Boltzmann simulations.  We identify a pattern-forming instability which produces an oscillating array of fluid vortices, and we elucidate which features govern the growth rate, wavelength, and saturation of the vortices. Our observation of pattern formation through odd mechanical response can inform models of biological patterning and guide engineering of odd dynamics in soft active matter systems.  
	\end{abstract}
	
	\maketitle
	
	A striking feature of non-equilibrium systems is their tendency to undergo spatiotemporal pattern formation \cite{cross1993pattern,cross2009pattern}.   Coherent structures such as convective rolls \cite{busse1978non}, Turing patterns \cite{turing1990chemical}, and pulsatile contractions of active gels \cite{staddon2022pulsatile} emerge spontaneously as the active driving in a system overcomes stabilizing dissipative forces.  Pattern-forming instabilities are biologically important since, for example, they are utilized by growing organisms for morphogenesis \cite{gross2017active}.  In many biological examples of soft active matter systems, patterns are driven by the interplay of an active contribution to the local stress \cite{marchetti2013hydrodynamics, simha2002hydrodynamic, doostmohammadi2018active} and a concentration field of chemical regulators \cite{bois2011pattern, kumar2014pulsatory, radszuweit2013intracellular, alonso2017mechanochemical, staddon2022pulsatile,del2022front}.  One can ask whether pattern formation in soft active matter systems can be reached through alternative routes which do not rely on active stresses and gradients of chemical regulators.  We show here that pattern formation in a viscoelastic fluid can occur without either of these features, provided that the system displays \textit{odd} non-equilibrium elastic responses to mechanical deformations.
	
	Odd elasticity, which complements the older theory of odd viscosity \cite{avron1998odd,souslov2019topological,soni2019odd,han2021fluctuating,liao2019mechanism,hargus2020time, epstein2020time}, has been developed by Vitelli and coworkers to describe elastic materials with internal energy-consuming degrees of freedom that do not obey several of the usual symmetries from classical elasticity theory \cite{scheibner2020odd, braverman2021topological, banerjee2021active, lier2022passive, fruchart2023odd}.  It has recently been reported that certain engineered and even biological systems exhibit odd elasticity: crystals of spinning magnetic colloids \cite{bililign2022motile} and starfish embryos \cite{tan2022odd}; certain active metamaterials \cite{chen2021realization}; and even muscle fibers \cite{shankar2022active} all transduce energy from an external or chemical drive into non-reciprocal pairwise interactions.  
	
	Whereas the predicted phenomenology of odd elastic systems, such as odd elastic waves and negative Poisson ratios \cite{scheibner2020odd, braverman2021topological}, has been appreciably mapped out, the full implications of odd responses in \textit{viscoelastic} materials remains to be explored. Some theoretical progress has been made in characterizing the thermodynamics and wave dispersion properties of odd viscoelastic materials \cite{scheibner2020odd, banerjee2021active, lier2022passive}.  These works identified novel transport properties and suggested ways that these properties could be experimentally detected; they also proposed that odd dynamics could be important for describing active biological materials like the actomyosin cortex.  However, exploring this possibility for complex models which capture the composite nature of biological active matter requires advances in simulation methods to allow for tensorial viscoelastic responses in the hydrodynamic description of multi-component active viscoelastic fluids.   
	
\begin{figure}[ht!]
		\begin{center}
			\includegraphics[width= \columnwidth]{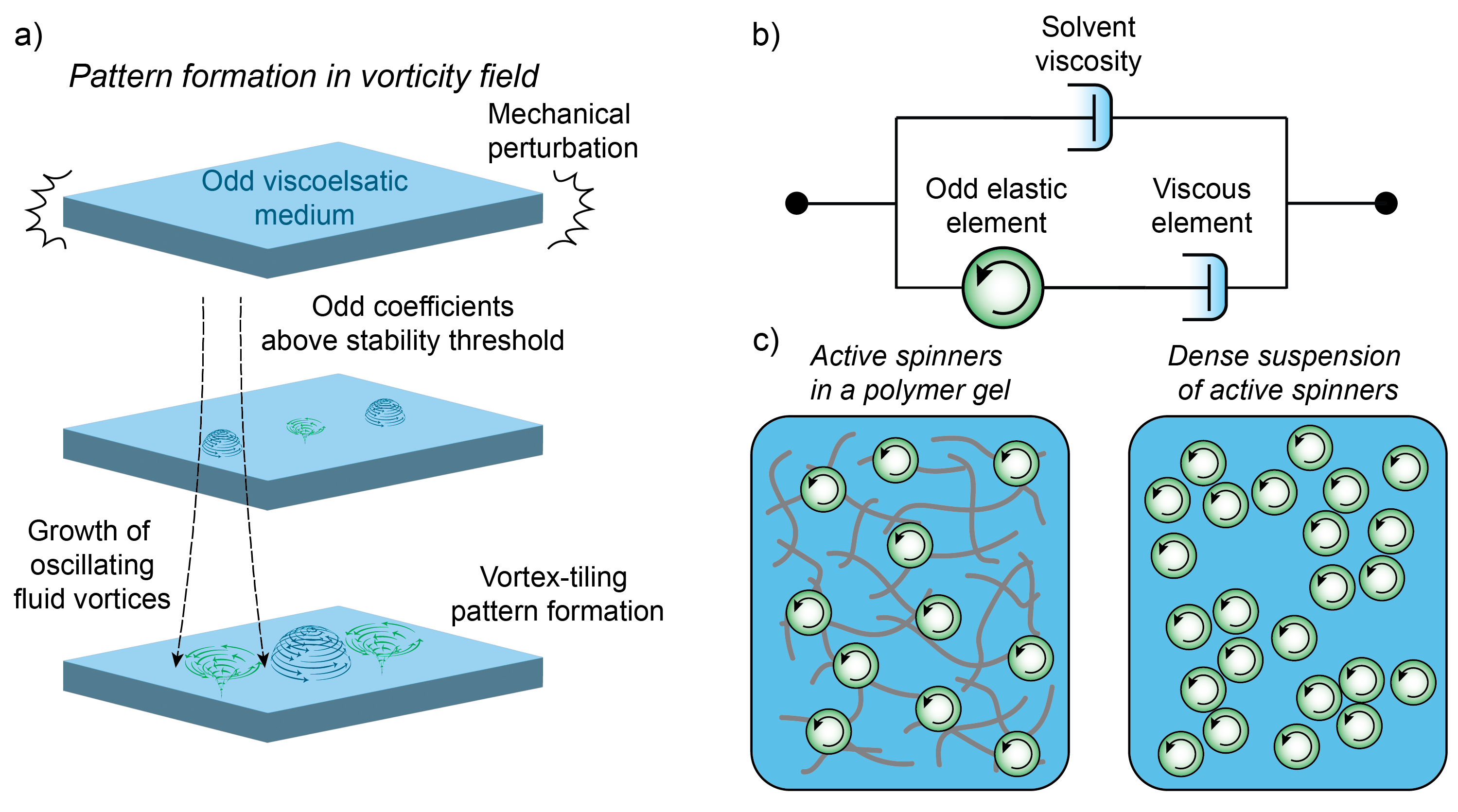}
			\caption{Odd viscoelastic fluids.  (a) Schematic illustration of the pattern formation instability observed in odd viscoelastic fluids.  (b) A three-element mechanical circuit, comprising a viscous solvent in parallel with an odd Maxwell element, represents a minimal model for an odd viscoelastic fluid.  (c) Two candidate systems which may display odd viscoelastic phenomenology.}
			\label{Schematic}
		\end{center}
	\end{figure}

	Here, we report on hydrodynamic simulations of a three-element active viscoelastic fluid using a recently developed extension of the hybrid lattice Boltzmann algorithm which can treat odd viscoelastic forces \cite{floyd2023simulating}. Combining these simulations with linear instability analysis, we demonstrate that the interaction of passive viscosity and active odd elasticity allows for the emergence of an oscillating vortex array with a tunable characteristic wavelength and growth rate, a feature not observed in previous simplified models of odd viscoelasticity (Figure \ref{Schematic}a).  We additionally show that the initial exponential growth of the vortices saturates if a shear-thickening non-linearity is included in the dynamics.  Our results suggest that such dynamical signatures may be generic to broad classes of odd viscoelastic systems encompassing various microscopic dynamics.  
	
	\begin{figure*}[ht!]
		\begin{center}
			\includegraphics[width= \textwidth]{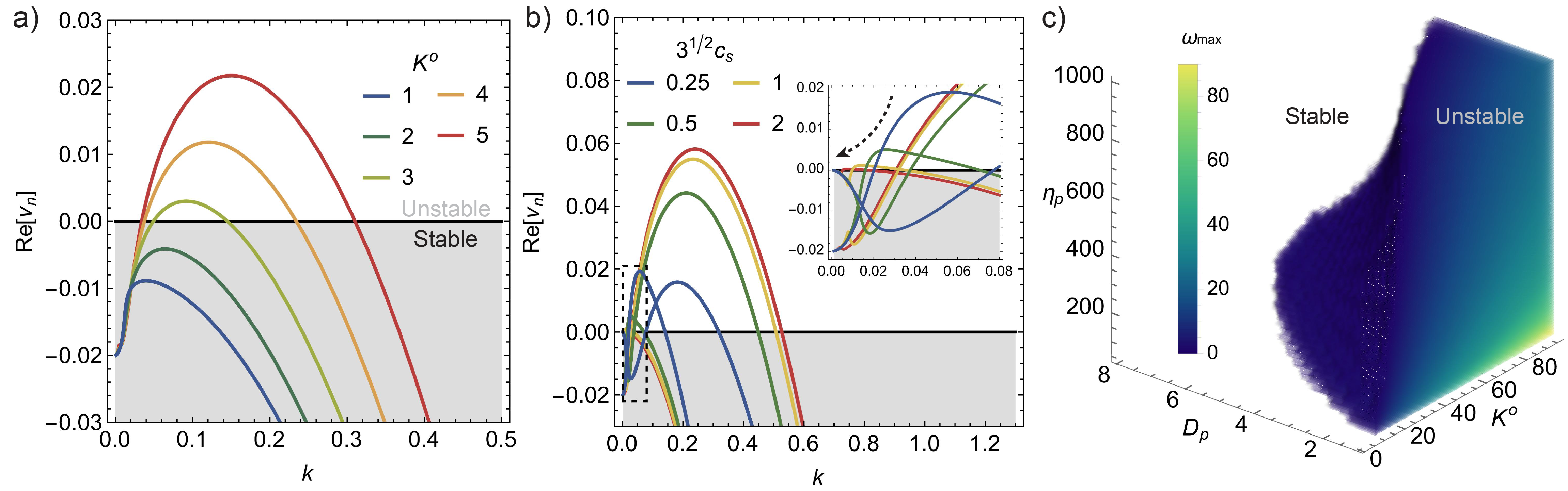}
			\caption{Dispersion relations.  To reduce the number of free parameters, all quantities in this figure are non-dimensionalized using the following physical scales: pressure $P = \mu$, length $L = \eta_\text{s} / \sqrt{\rho_\text{s} \mu}$, and flow speed $V = \sqrt{\mu / \rho_\text{s}}$ where $\rho_\text{s}$ is the density of the fluid in the uniform state.    The default parameters are given in the Supplementary Material. (a) The real part of one of the nine branches $\nu_n$ is shown as $K^o$ is varied.  (b)  Two branches are shown as $c_\text{s}$ is varied.  The boxed region is blown up and displayed as an inset to allow easier visualization.  The dashed arrow indicates that as the system becomes less compressible with increasing $c_\text{s}$, the compressibility-dependent branch of the unstable region decreases to zero.  (c)  Three-dimensional heatmap of the maximum real part of the growth rate $\omega_\text{max}$ in the incompressible limit.   
			}
			\label{DispersionRelationsMain}
		\end{center}
	\end{figure*}
	
	\textit{Odd viscoelastic fluid model}.
	Our model for odd viscoelasticity in this paper is an \textit{odd Jeffrey fluid}.  The usual Jeffrey fluid consists of a solvent phase in which a viscoelastic Maxwell material is immersed \cite{bird1987dynamics, larson2013constitutive}, and it has been identified as a good description of biological systems like the cytoplasm \cite{Xiee2115593119, najafi2023size}.  In our case, while we treat the viscosities of the solvent and viscoelastic phases as scalar, we treat the elastic contribution to the fluid stress using the theory of odd elasticity.  The mechanical circuit describing this viscoelastic model is depicted in Figure \ref{Schematic}b.  It can be shown that this model can map directly onto other three-element viscoelastic fluid models, such as one in which the solvent viscosity acts in series with a Kelvin-Voigt element; see Refs.\ \citenum{bird1987dynamics, najafi2023size}.  We expect that an odd Jeffrey fluid could be physically realized in at least two types of systems: one in which active spinners are linked together through a polymer network \cite{howard2019structure}, and one in which attractive interactions between the active spinners cause them to form a dense suspension through viscoelastic phase separation \cite{tanaka2000viscoelastic,patrick2008direct} (Figure \ref{Schematic}c).  A key feature of these systems is that the spinners are not confined to a crystalline order, which would require description as an odd elastic or viscoelastic \textit{solid} \cite{scheibner2020odd,bililign2022motile,tan2022odd,petroff2015fast}.

	The dynamical equations governing the evolution of the odd Jeffrey fluid are
	\begin{align}
		\partial_t \rho &= - \partial_i (\rho v_i) \label{eqdenscons} \\
		p &= c_\text{s}^2 \rho \label{eqeos} \\
		\rho D_t v_i &= -\partial_i p + 2 \eta_\text{s} \partial_k \Psi_{ik} + \partial_k \sigma_{ik}^\text{p} + f_i  \label{eqNavStokes} \\
		\mathcal{D}_t\sigma_{ij}^\text{p} &= C_{ijkl}\partial_{k}v_l - \eta_\text{p}^{-1}C_{ijkl}\sigma_{kl}^\text{p} + D_\text{p} \partial_{kk}\sigma_{ij}^\text{p}. \label{eqMaxwell}
	\end{align}
	Here, $\rho$ is the fluid density, $\mathbf{v}$ is its velocity, $p$ is the pressure, $c_\text{s}$ is the speed of sound in the fluid, $\eta_\text{s}$ is the solvent's dynamic viscosity, $\Psi_{ij} \equiv \left(\partial_i v_j + \partial_j v_i\right)/2$ is the symmetric strain rate tensor, and $\boldsymbol{\sigma}^\text{p}$ is the viscoelastic contribution to the stress tensor.  \color{\editcolor} The isothermal equation of state, Equation \ref{eqeos}, implies that the fluid is weakly compressible \cite{kruger2017lattice}; see Ref.\ \citenum{lier2023lift} for recent work on the interplay of weak compressibility and odd viscous forces. \color{black}  The term $  \mathbf{f} $ in the Navier-Stokes equation is an optional external force field. $\mathbf{C}$ is a rank four odd elasticity modulus tensor, $\eta_\text{p}$ is the dynamic viscosity of the viscoelastic phase (assumed to be scalar here), and $D_\text{p}$ is the viscoelastic stress diffusion constant \cite{olmsted2000johnson}.  Odd tensorial viscosity of the viscoelastic phase could be straightforwardly incorporated in this model by generalizing the coefficient $\eta_\text{p}^{-1}C_{ijkl}$ in Equation \ref{eqMaxwell} as a rank four relaxation tensor $R_{ijkl}$ \cite{banerjee2021active}.  This would not affect the functional form of the model, so we omit this for simplicity. Further, $\partial_t$ is the partial derivative with respect to time, ${D_t \equiv \partial_t + v_k \partial_k}$ is the material derivative, and ${\mathcal{D}_t X_{ij} \equiv D_t X_{ij} + \Omega_{ik}X_{kj} - X_{ik}\Omega_{kj}}$ is the corotational derivative of the tensor $\mathbf{X}$, with the vorticity tensor defined as ${\Omega_{ij} \equiv \left(\partial_i v_j - \partial_j v_i\right)}/2$.  If the upper convected derivative were used instead of the corotational derivative in Equation \ref{eqMaxwell}, we would have the Oldroyd-B model.  In the subsequent linear instability calculation, however, these two derivatives are equivalent because they both reduce to $\partial_t$ to linear order, and our analytical results thus hold for the Oldroyd-B model as well.

	In this work we consider an isotropic odd elastic modulus tensor $C_{ijkl}$, whose form was derived in Ref.\ \citenum{scheibner2020odd}:
	\begin{align}
		C_{ijkl} = &\ B \delta_{ij}\delta_{kl} + \mu\left(\delta_{il}\delta_{jk} + \delta_{ik}\delta_{jl} - \delta_{ij}\delta_{kl}\right) \nonumber \\
		&+ K^o E_{ijkl} - A \epsilon_{ij}\delta_{kl},
		\label{Cdef}
	\end{align}
	where $\delta_{ij}$ is the Kronecker delta, $\epsilon_{ij}$ is the Levi-Civita tensor, and $E_{ijkl} \equiv \left(\epsilon_{ik}\delta_{jl} + \epsilon_{il}\delta_{jk} + \epsilon_{jk}\delta_{il} + \epsilon_{jl}\delta_{ik} \right)/2$. The bulk ($B$) and shear ($\mu$) moduli are found in classical elasticity theory, but the modulus $A$, which transforms a dilatational deformation into torque (but not vice versa), and $K^o$, which antisymmetrically couples the two shear modes, are the active ``odd'' moduli.  \color{\editcolor} Equation \ref{eqMaxwell} is a phenomenological generalization of a standard Maxwell material to include odd elastic coefficients. It does not correspond to a specific microscopic system, instead serving as a general model to explore the repercussions of non-reciprocity in a composite viscoelastic material.  In the Supplementary Methods, however, we illustrate how one can coarse-grain a microscopic ``non-reciprocal elastic dumbbell'' model to yield continuum equations with emergent odd coefficients like $K^o$.  \color{black}

	\textit{Results}.
	The stability of the homogeneous state of the odd Jeffrey fluid is controlled by an intricate balance between stabilizing and destabilizing forces, the relative magnitudes of which depend on the parameters entering Equations \ref{eqdenscons}-\ref{Cdef}.  In Supplementary Methods Section IIA we derive the dispersion relation $\nu(k)$ for the growth of plane wave perturbations with the ansatz $ e^{\nu(k) t}e^{i \mathbf{k}\cdot\mathbf{r}}$ for these dynamics.  Linear instabilities occur when $\omega(k) > 0$ for some wavenumber $k$, where ${\omega(k) \equiv \text{max}_n \text{Re}[\nu_n(k)]}$ is the largest of the real part of nine branches of the dispersion relation $\nu(k)$.  The dispersion relation is complicated but reduces to the linear form derived in Ref.\ \citenum{banerjee2020actin} in the special case ${\eta_\text{s} =0, \ \eta_\text{p} \rightarrow \infty, \ D_\text{p} = 0}$, and $A = 0$.  \color{\editcolor}  Although the stabilizing forces in this composite viscoelastic fluid are more complex than those in a one-component viscoelastic solid as considered in Ref.\ \citenum{scheibner2020odd}, the intuition provided there of odd work cycles driving active waves also applies to understand the instabilities found in our model. \color{black}
	
	We studied how the various parameters control the system's stability by plotting for each parameter the dispersion relation ${\omega_\text{max} \equiv \text{max}_k  \omega(k) }$ over a range of parameter values (Figure \ref{DispersionRelationsMain}a; see Supplementary Figure 1 for several other parameters).  Key drivers of the instability include the odd moduli $A$ and $K^o$: when their values lie outside a threshold set by the remaining parameters, the homogeneous state is unstable.  Furthermore, $K^o$ alone is sufficient to cause instability, and $A$ cannot cause instability if $K^o = 0$.  The two parameters work cooperatively if their signs agree, such that if $K^o > 0$ then the instability growth rate increases as $A$ increases, but if $K^o < 0$ the growth rate increases as $A$ decreases (Supplementary Figure 2).  We also find that the value of the fastest growing wavenumber ${k_\text{max} \equiv \text{argmax}_k \omega(k)}$ increases with $K^o$ (Figure \ref{DispersionRelationsMain}a and Supplementary Figure 4a) and either increases or decreases with $A$ depending on the relative signs of $K^o$ and $A$.  
	
	\begin{figure}[ht!]
		\begin{center}
			\includegraphics[width= \columnwidth]{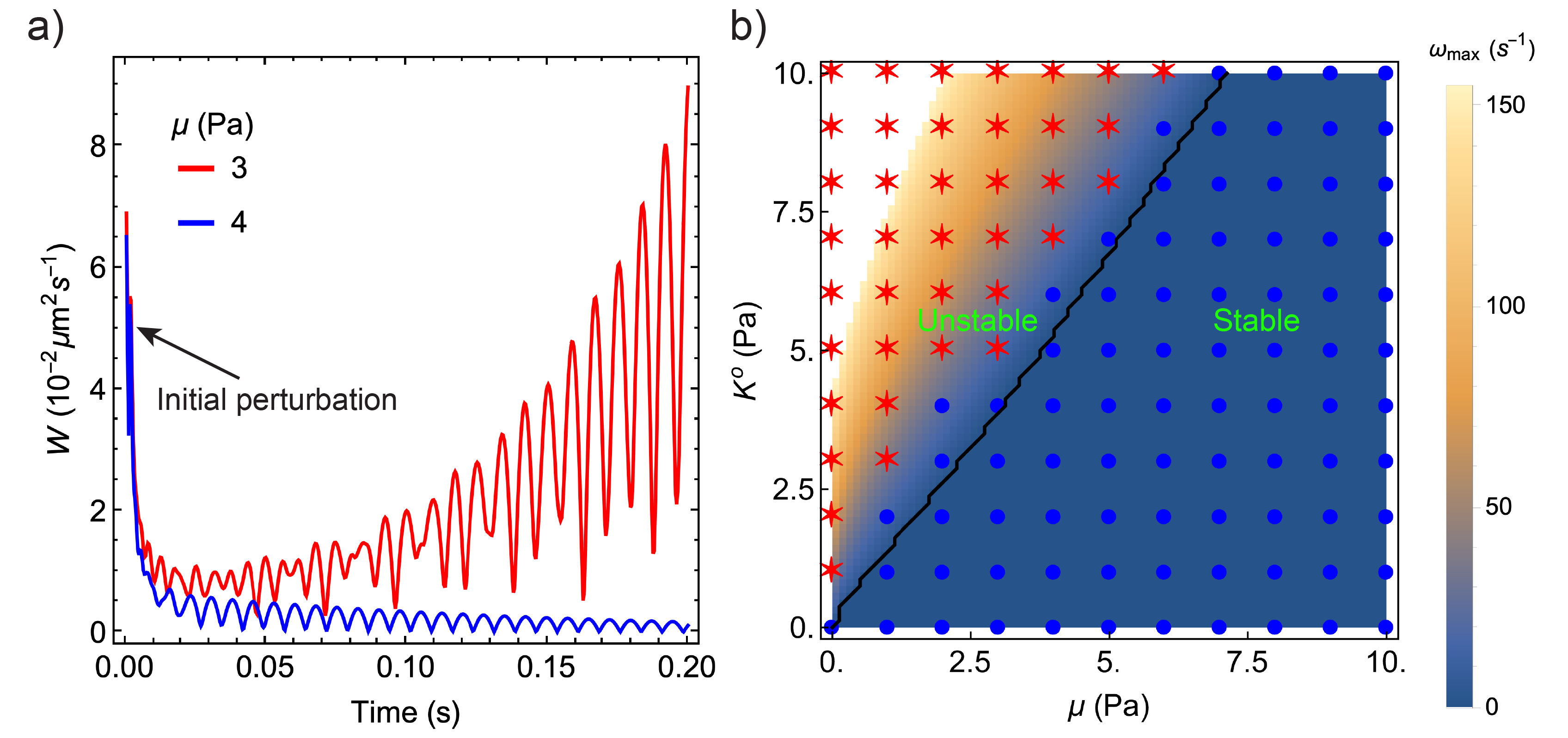}
			\caption{Numerical validation of instability threshold.  (a) For ${K^o = 5 \ \text{Pa}}$, trajectories of $W$ for two values of $\mu$.  See Supplemental Movies for videos of these simulations.  (b) A heatmap of ${\text{max}\{0, \omega_\text{max} \}}$, with a contour at $\omega_\text{max} = 0$ drawn in black.  When $\omega_\text{max} > 0$, the homogeneous state of the system is unstable.  The symbols are simulation data, with red stars representing a detected instability for that condition and blue circles representing a lack of detected instability.}
			\label{MainAgreement}
		\end{center}
	\end{figure}
	
	The nature of the instability threshold qualitatively changes in the incompressible limit $c_\text{s} \rightarrow \infty$ (where dilatational deformations disappear, i.e., $\partial_k v_k = 0$).  First, as one might expect, the instability no longer depends on the moduli $A$ or $B$ which couple dilatational deformations to, respectively, a torque and an isotropic stress.  Additionally, in the compressible case we typically observe two branches of the dispersion relation $\nu_n(k)$ which can take on real positive values for some $k$.  However, in the incompressible limit one of these branches shrinks below $0$ and remains stable for all $k$ (Figure \ref{DispersionRelationsMain}b). 
	
	The shear modulus $\mu$, the viscosities $\eta_\text{s}$ and $\eta_\text{p}$, and the stress diffusion constant $D_\text{p}$ have predominantly stabilizing effects, causing $\omega(k)$ to decrease as their values increase (Figure \ref{DispersionRelationsMain}c and Supplementary Figure 1).  We note that $\eta_\text{s}$ is a key parameter which suppresses the linear relationship $\omega(k) \propto k$ at large $k$ (Supplementary Figure 3).  This allows for a finite $k_\text{max}$ and thus a finite length scale of the instability.  In previous work \cite{banerjee2021active} $\eta_\text{s}$ was set to zero, precluding the observation of pattern formation since all wavelengths are unstable if $\omega(k) \propto k$.

	\begin{figure*}[ht!]
		\begin{center}
			\includegraphics[width= \textwidth]{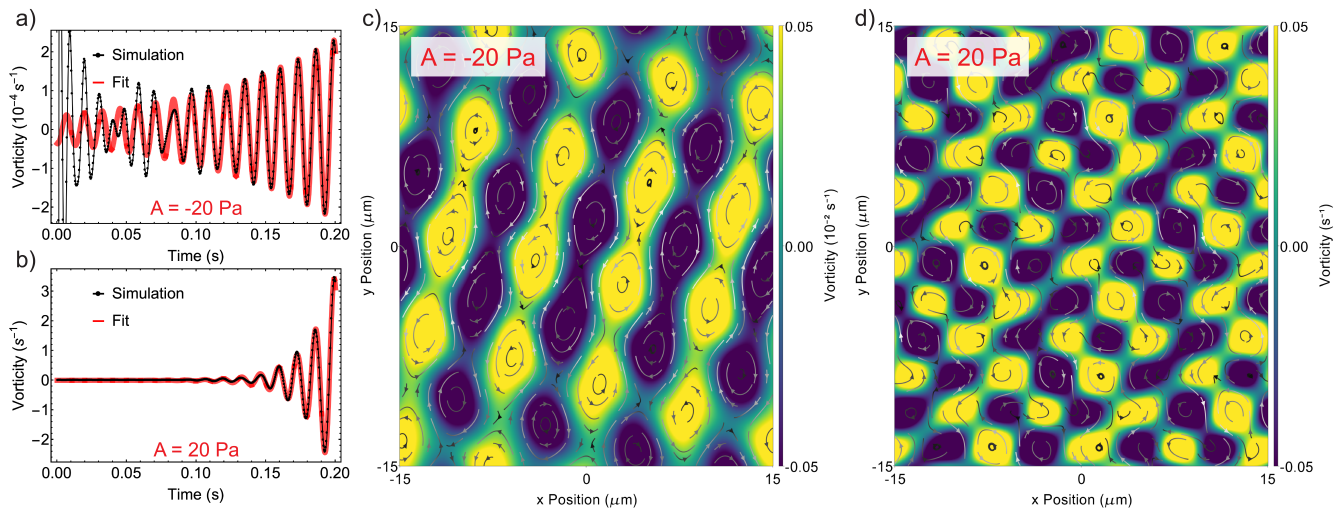}
			\caption{Simulations of pattern formation.  The default parameters are given in the Supplementary Material.  (a) The vorticity over time at the point $(-5 \ \mu \text{m}, -5 \ \mu \text{m})$ is shown as black symbols for ${A = -20 \ \text{Pa}}$. The red curve is the fit of Equation \ref{omegafunc} to these data points.  The fitting parameters $b$ and $c$ are $9 \ \text{s}^{-1}$ and $490 \ \text{s}^{-1}$ .  (b)  The same as panel a, but for ${A = 20 \ \text{Pa}}$. The fitting parameters $b$ and $c$ are $51 \ \text{s}^{-1}$ and $473 \ \text{s}^{-1}$.  (c) A simulation snapshot at ${t = 0.13 \ \text{s}}$ for the condition ${A = -20 \ \text{Pa}}$.  Color represents vorticity $\Omega$ and the streamlines range from black to white as the represented velocity increases.  (d) The same as panel c, but for ${A = 20 \ \text{Pa}}$.  See Supplemental Movies for videos of these simulations.
			}
			\label{Composite}
		\end{center}
	\end{figure*}
	
	We next sought to study the growth of the instability in the compressible case using lattice Boltzmann simulations.  To simulate an odd Jeffrey fluid, we apply a recently developed implementation of the hybrid lattice Boltzmann algorithm \cite{floyd2023simulating}.  To excite the instability in simulation we apply a short, periodic, random force $\mathbf{f}(\mathbf{r})$ (see Supplementary Methods Section IIB) and then evolve the system.  As a readout of the instability, we use the total absolute vorticity in the system ${W \equiv \int |\Omega(\mathbf{r})| d\mathbf{r}}$, where ${\Omega \equiv \partial_x v_y - \partial_y v_x = 2 \Omega_{xy}}$.  Above the instability threshold, $W$ oscillates and grows exponentially in time (Figure \ref{MainAgreement}a). If the growth rate is fast enough that $W$ exceeds the value it attained during the initial perturbation in $0.2 \ \text{s}$, we conclude that the fluid is unstable.  In Figure \ref{MainAgreement}b we show that the conditions of $K^o$ and $\mu$ which are predicted to be unstable from the linear instability calculation are matched by those which produce a detected instability in simulation.   In Supplementary Figure 4b we also show that the fastest growing wavelength of the instability matches the characteristic wavelength detected in simulation.  
	
	The spatial structure of the pattern is a regular periodic array of vortices with alternating handedness.  The vorticity at a given point oscillates and grows exponentially in time, as shown in Figures \ref{Composite}a,b, where the vorticity at a point is fit to the functional form
	\begin{equation}
		\Omega(t) = a \exp(b t)\cos(c t + d). \label{omegafunc}
	\end{equation}
	We observe checkerboard and striped patterns as illustrated in Figure \ref{Composite}c,d and in Supplementary Figure 6.  The periodic patterns do not travel but instead resemble standing waves.  This instability falls in type $\text{I}_\text{o}$ of the classification of Cross and Hohenberg \ \cite{cross1993pattern}, being periodic in space and oscillatory in time.  Although we have focused on the vorticity $\Omega(\mathbf{r})$ as the pattern-forming field, we note that patterning appears for other fields as well, including the divergence, density, and torque, as shown in Supplementary Figure 7.
	
	The initial exponential growth of the instability can in principle saturate due to various nonlinearities.  The advective term in the Navier-Stokes equation, which is neglected for unsteady Stokes flows at low Reynolds number, is one possibility.  We account for this term in our simulations, but we typically observe that the lattice Boltzmann algorithm becomes numerically unstable due to large fluid velocities before saturation from this term occurs.  Another possible source is the nonlinear correction to the elastic forces experienced for large deformations, which we neglect here.  Treating odd effects in the framework of finite elasticity requires additional theoretical development.  Instead, we study here saturation caused by a shear-thickening nonlinearity which can result, for instance, from flocculation of dilatant viscoelastic suspensions like blood \cite{boersma1990shear, bodnar2009numerical}.  We consider a Carreau form \cite{carreau2021rheology}
	$
	\eta_\text{p}(\Psi_{ij}) = \eta_\text{p}^0\left(1 + 2 \beta^2 \Psi_{ij} \Psi_{ij} \right)^{(n-1)/2}
	$
	which we use in Equation \ref{eqMaxwell}.  The parameter $\beta$ sets the scale at which the shear flow $\Psi_{ij}$ begins to alter the viscosity, and the exponent $n$ determines if the system is shear-thinning ($n<1$) or thickening ($n>1$); here we use $n = 1.5$.  In Supplementary Figure 5 we display the trajectory $W(t)$ for several values of $\beta$, showing that this nonlinearity can significantly tune the flow rate in the pattern forming state of an odd viscoelastic fluid.

	\textit{Conclusion}.
	We have shown that ``odd" moduli can provide a mechanism for pattern formation in non-equilibrium viscoelastic fluids.  Whereas in typical soft active matter systems pattern formation is driven by active stresses and chemical regulators \cite{simha2002hydrodynamic, doostmohammadi2018active,bois2011pattern, kumar2014pulsatory, radszuweit2013intracellular, alonso2017mechanochemical, staddon2022pulsatile}, here it is driven by active elastic response to mechanical deformations.  Given that pattern formation and wave propagation due to active stresses can template developmental processes \cite{gross2017active}, our discovery of another mechanism for pattern formation may have biological implications.  Odd elastic forces could also interact with active stresses and chemical regulators.  This may introduce new features to current models of traveling waves, pulsatile motions, and other dynamical patterns known to occur in biological or bio-inspired materials like actomyosin sheets \cite{bois2011pattern, staddon2022pulsatile, banerjee2017actomyosin, del2022front}.
	
	\color{\editcolor}  Collectives of rollers \cite{han2020emergence, han2020reconfigurable, zhang2022polar, han2023globally} as well as both reconstituted and \textit{in vivo} cytoskeletal systems \cite{tee2015cellular, schaller2010polar} exhibit chiral and vortical flows similar to those reported here.  While models for these systems are not currently framed using the theory of odd viscoelasticity, it should be possible to construct emergent, coarse-grained descriptions of their dynamics in terms of odd coefficients.  In Supplementary Methods we provide an example of this type of coarse-graining for a microscopic ``non-reciprocal elastic dumbbell'' model; this derivation recapitulates the key coefficient $K^o$ driving instabilities in our phenomenological dynamical equations.  We note that coarse-graining cytoskseletal systems poses a challenge that the constituent force dipoles are anisotropic, in contrast with the current isotropic model of odd elasticity.
	\color{black}
	
	We focused here on the linear instability of an odd Jeffrey fluid, but future work could clarify its rheological and dynamical properties.  Detectable signatures of odd dynamics should be present even below the instability threshold.  For example, we expect that in canonical setups such as Couette or Pouseille flow of a compressible odd Jeffrey fluid, one may find transverse components of the flow, analogous to the Hall effect.  \color{\editcolor}  A recent theoretical study clarifies the expected dynamics experienced by a probe particle immersed in an odd viscoelastic fluid \cite{duclut2023probe}.  \color{black}It would also be worth exploring whether features of pattern formation in other active systems such as screening by substrate friction \cite{doostmohammadi2016stabilization}, wavelength selection by confinement \cite{chandrakar2020confinement}, and transitions to turbulence \color{\editcolor}\cite{wu2017transition, datta2022perspectives, de2023pattern} \color{black} occur in odd viscoelastic fluids.

	\section*{Acknowledgments}
	We wish to thank Vincenzo Vitelli and his group for helpful discussions.  This work was mainly supported by funds from DOE BES Grant DE-SC0019765 (CF and SV). ARD acknowledges support from the University of Chicago Materials Research Science and Engineering Center, which is funded by the National Science Foundation under award number DMR-2011854. CF acknowledges support from the University of Chicago through a Chicago Center for Theoretical Chemistry Fellowship.  The authors acknowledge the University of Chicago’s Research Computing Center for computing resources.

\clearpage
\pagebreak
\onecolumngrid
\tableofcontents

\singlespacing

\section{Supplementary Figures}

\begin{figure}[h!]
	\begin{center}
		\includegraphics[width=\textwidth]{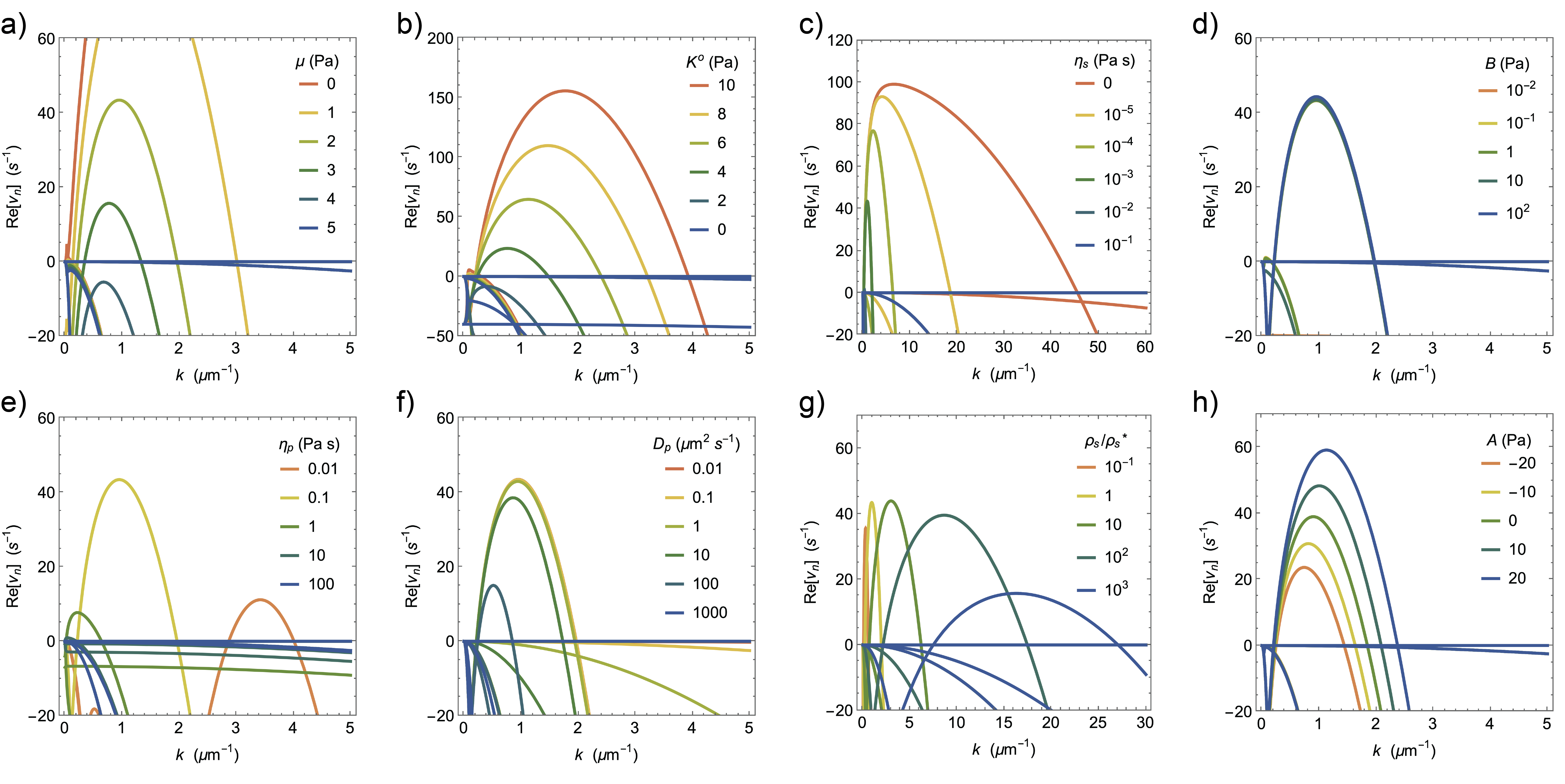}
		\caption{The dispersion relations as the indicated parameters are varied.  The full set of nine solutions $\text{Re}[\nu_n(k)]$ to the dispersion relationship are shown for each condition.  The default parameters are listed in Table \ref{VEParamsInst} below.  In panel (g), $\rho_\text{s}^* = 2\times10^7 \ \text{kg/m}$$^3$.}
		\label{DispersionRelations}
	\end{center}
\end{figure}

\begin{figure}[h!]
	\begin{center}
		\includegraphics[width=8 cm]{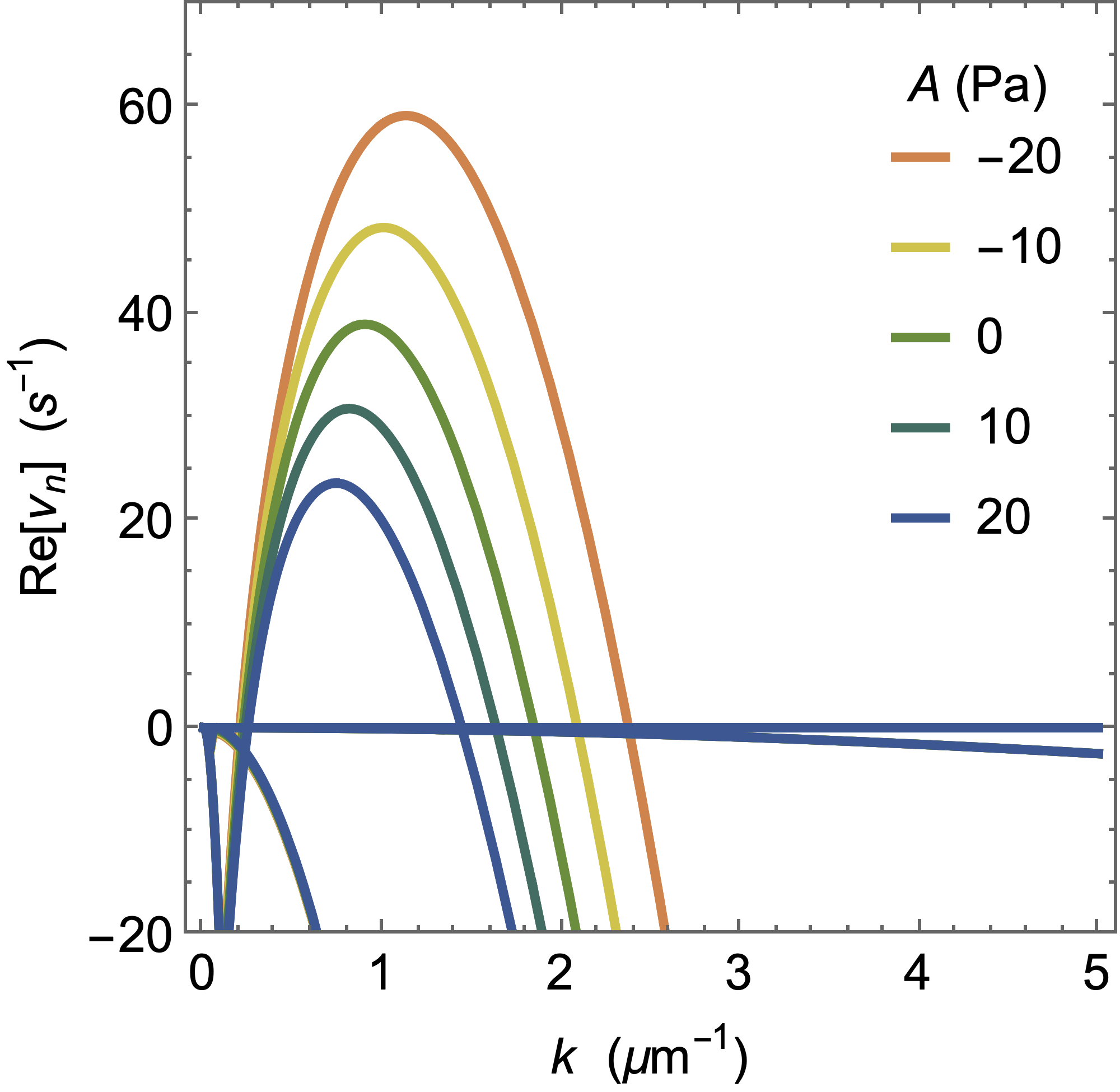}
		\caption{The solutions $\text{Re}[\nu_n(k)]$  as $A$ is varied, with $K^o = - 5 \ \text{Pa}$.  See Figure \ref{DispersionRelations}h for the corresponding plot when $K^o = 5 \ \text{Pa}$.}
		\label{DispersionRelationsAnK}
	\end{center}
\end{figure}

\begin{figure}[h!]
	\begin{center}
		\includegraphics[width=8 cm]{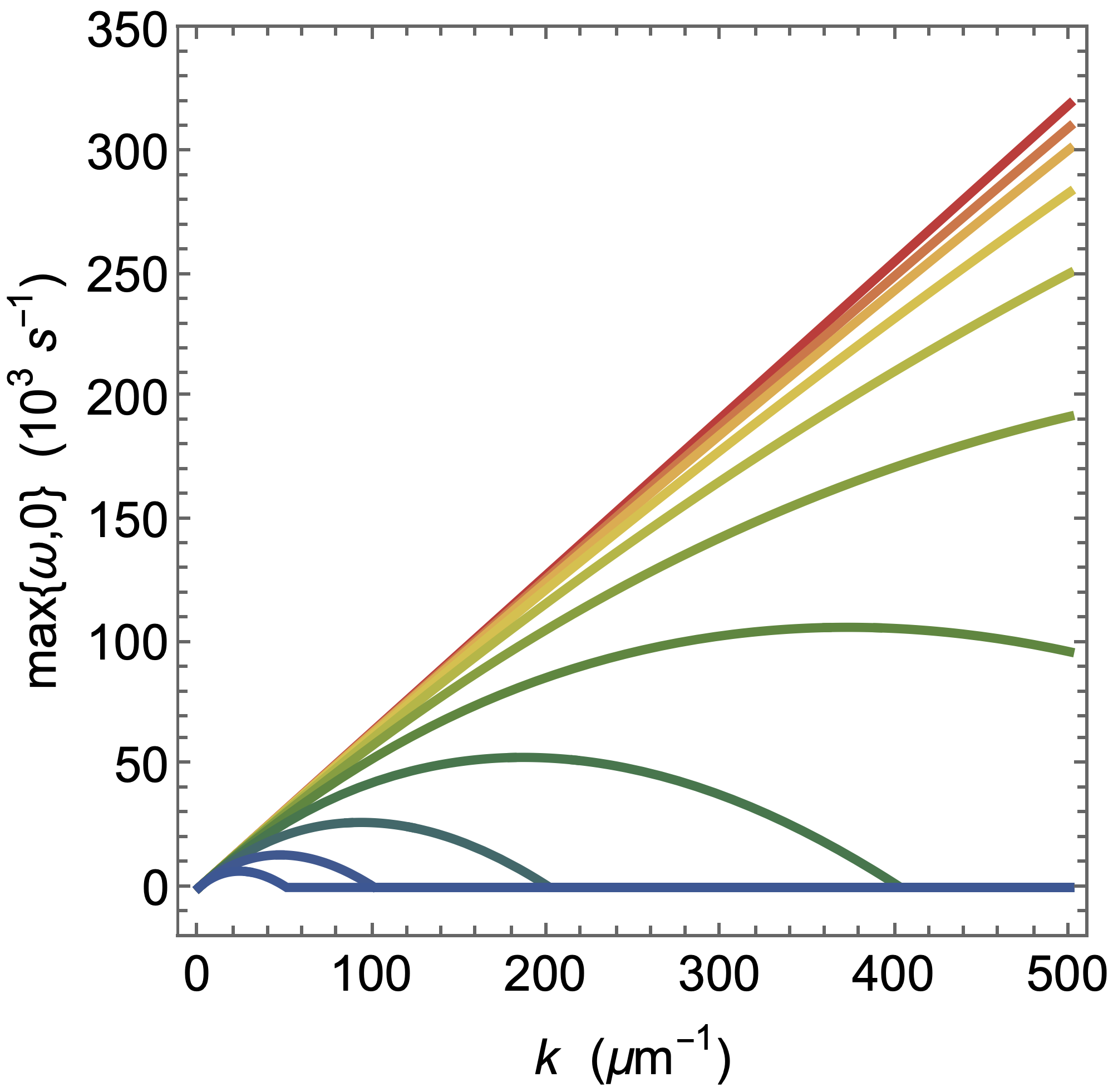}
		\caption{The function $\text{max}\{\omega, 0\}$ as $\eta_\text{s}$ is varied in the special case considered in Ref.\ \citenum{banerjee2021active}, with $K^o = 100 \ \text{Pa}$, $A = 0 \ \text{Pa}$, $\eta_\text{p} \rightarrow \infty \ \text{Pa s}$, and $D_\text{p} = 0 \ \mu\text{m}^2 / \text{s}$.  The remaining parameters are listed in Table \ref{VEParamsInst}. The red line corresponds to $\eta_\text{s} = 0 \ \text{Pa s}$, and the colors range from red to blue as $\eta_\text{s}$ increases by powers of $2$ from  $10^{-6}$ to $5.12 \times 10^{-4} \ \text{Pa s}$.}
		\label{EtasLinearLimit}
	\end{center}
\end{figure}

\begin{figure}[h!]
	\begin{center}
		\includegraphics[width=\textwidth]{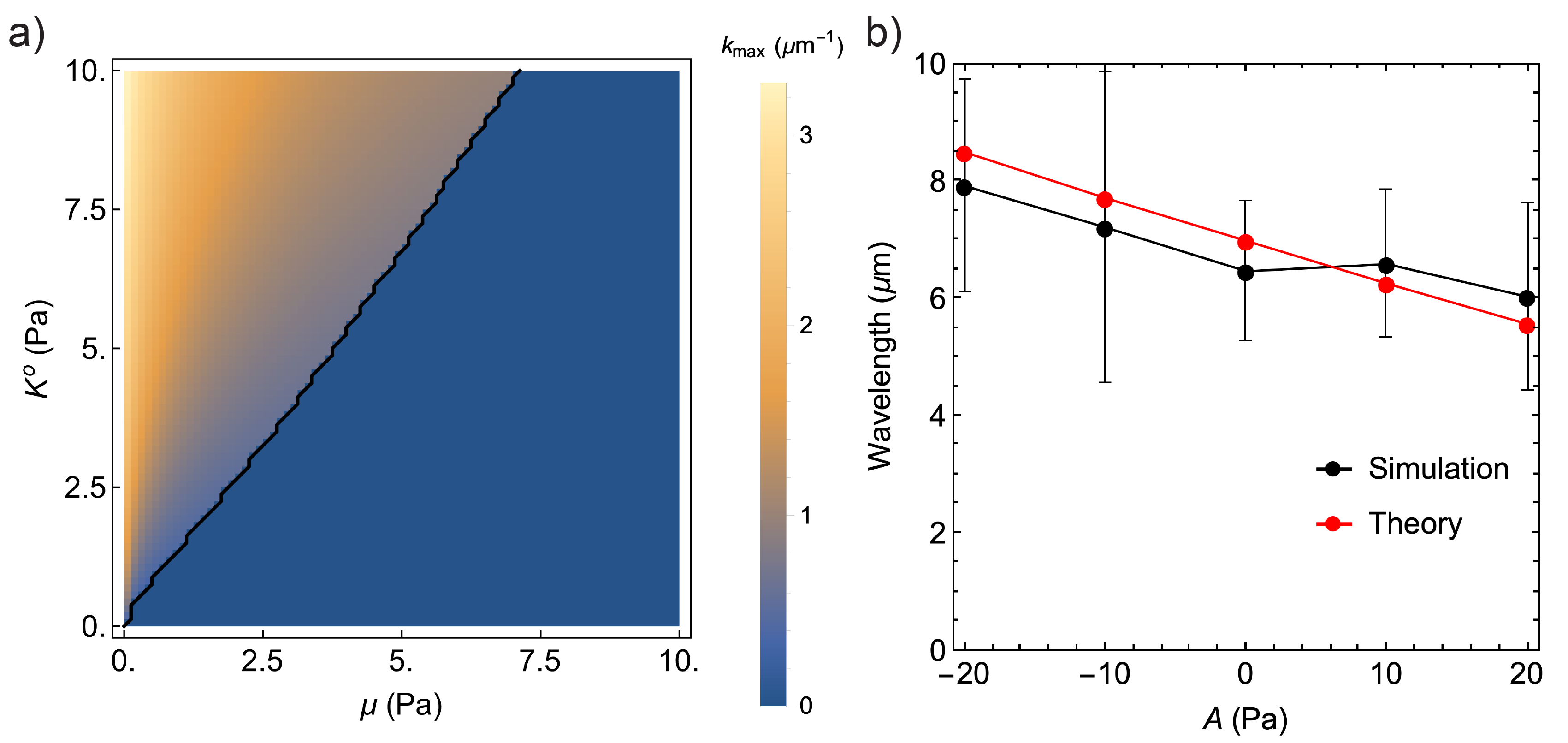}
		\caption{(a) A heatmap of $\text{max}\{0, k_\text{max} \}$ as $K^o$ and $\mu$ are varied, with a contour at ${\omega_\text{max} = 0}$ drawn in black.  (b) The spatial wavelengths from theory (red points, computed as $2\pi / k_\text{max}$) and simulation (black points, see Supplementary Methods for details) as $A$ is varied.  The error bars represent the standard deviation over the time interval $[0.13\ \text{s}, 0.16\ \text{s}]$ and $5$ trials.  We measured the characteristic wavelength of the patterns while varying $A$ by first finding the spatial correlation function of the vorticity field $C_\Omega(r)$, which is oscillatory due to the patterning, and then using the position of the first minimum of $C_\Omega(r)$ as an estimate of half the wavelength (see Supplementary Methods Section IIB for more details).  We find that for a given condition the characteristic wavelength has large variations over multiple random initial perturbations (see Supplementary Figure 5), yet it tracks the trends predicted from the analytical theory on average.  
		}
		\label{SIComposite}
	\end{center}
\end{figure}

\begin{figure}[ht!]
	\begin{center}
		\includegraphics[width= 8cm ]{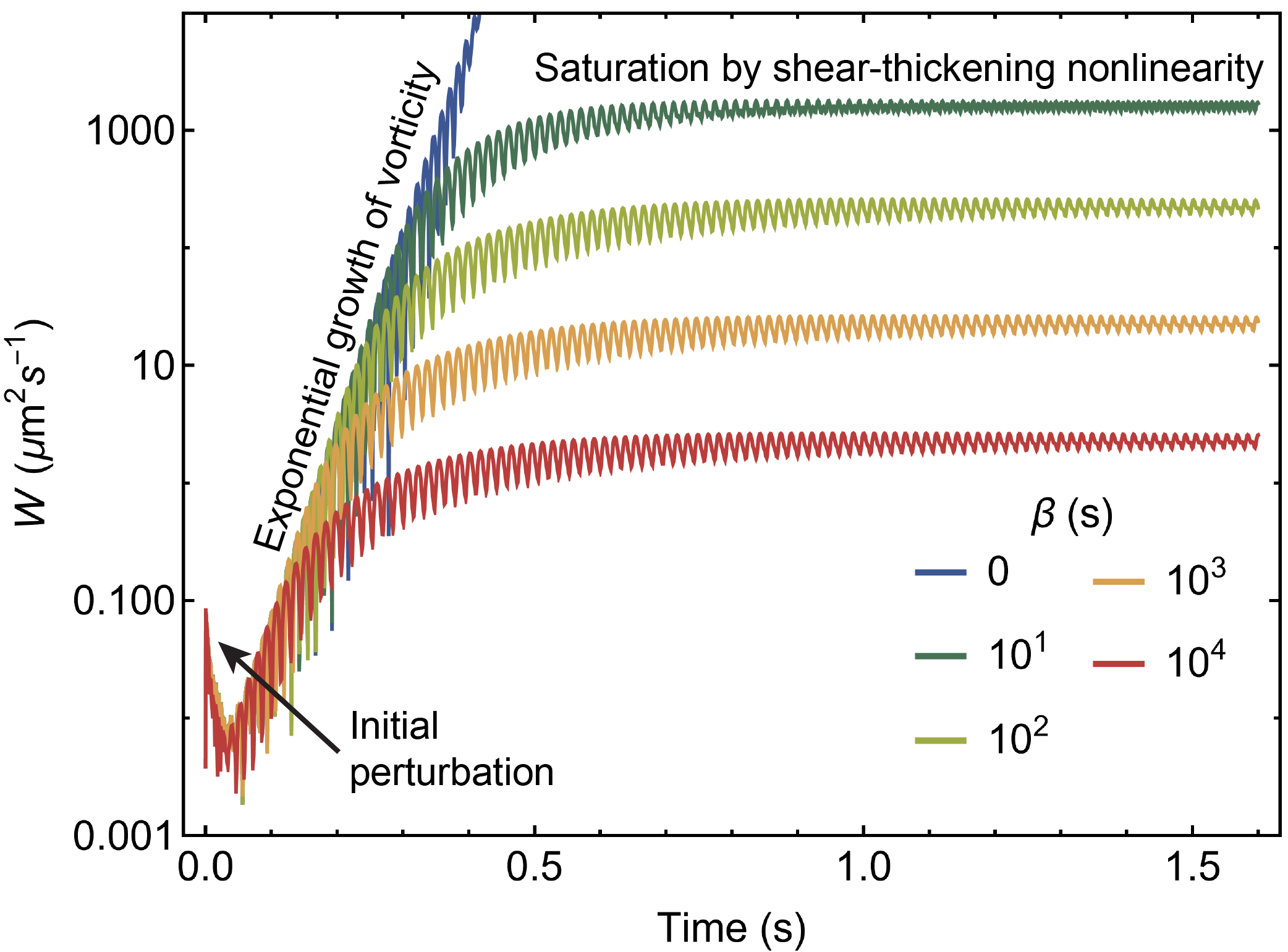}
		\caption{Saturation of growth by a shear-thickening nonlinearity.  The total absolute vorticity $W$ is plotted against time as the parameter $\beta$ of the Carreau form for shear-thickening is varied.  The same random seed was used for the initial perturbation in each of these simulations.  Here $\eta_p^0 = 0.1 \ \text{Pa s}$ and $N_x = Ny = 250$, and the remaining parameters are given in the Supplementary Material. }
		\label{betaSatAn}
	\end{center}
\end{figure}

\begin{figure}[h!]
	\begin{center}
		\includegraphics[width=\textwidth]{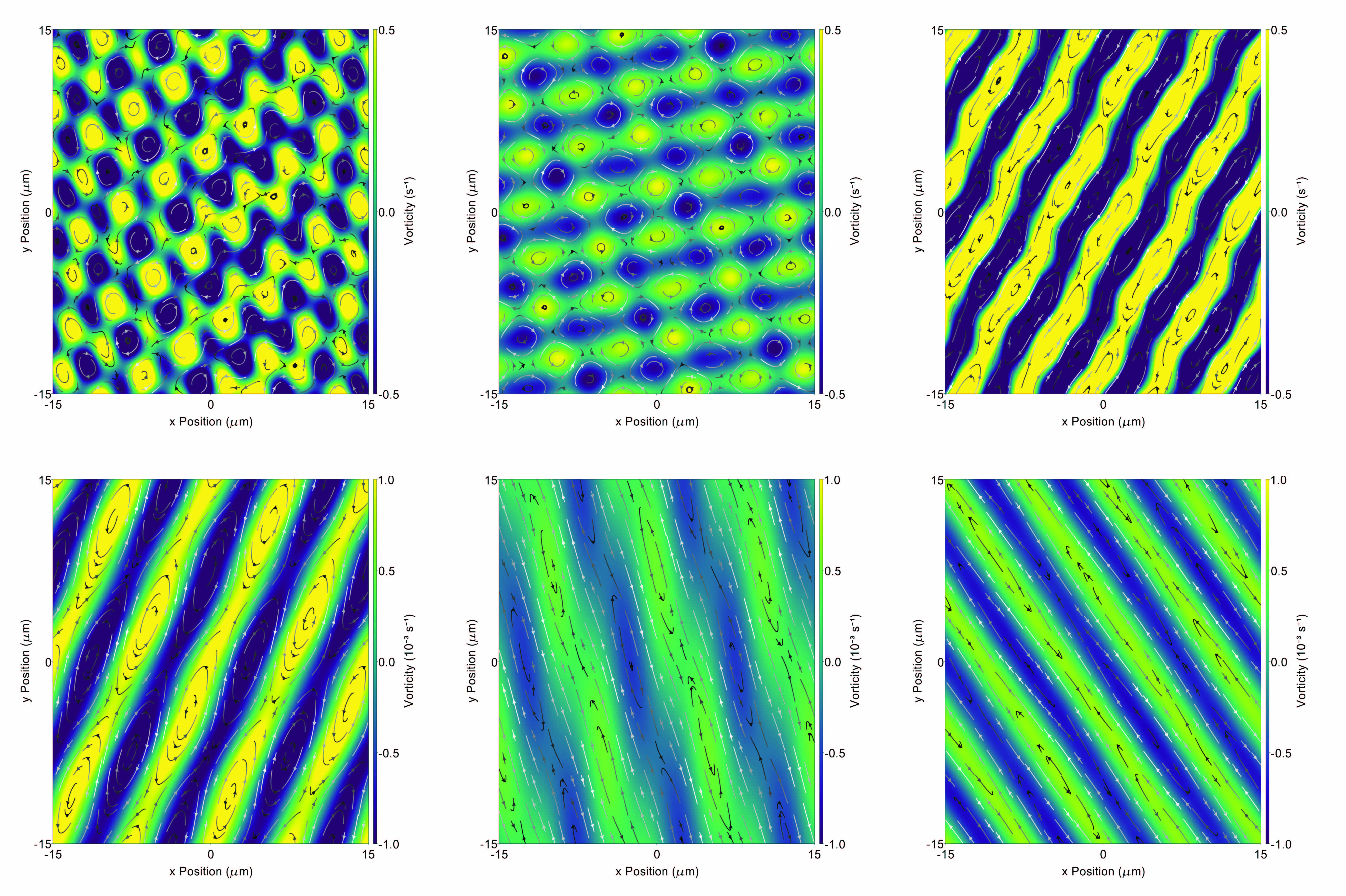}
		\caption{Simulation snapshots of the vorticity $\Omega$ at $t = 0.16 \ \text{s}$ for repeated trials of the condition $A = 20 \ \text{Pa}$ (top row) and $A = -20 \ \text{Pa}$ (bottom row).  Each simulation was initially perturbed using a different random force field, giving rise to the observed variability in patterns.}
		\label{MultipleTrials}
	\end{center}
\end{figure}

\begin{figure}
	\begin{center}
		\includegraphics[width=\textwidth]{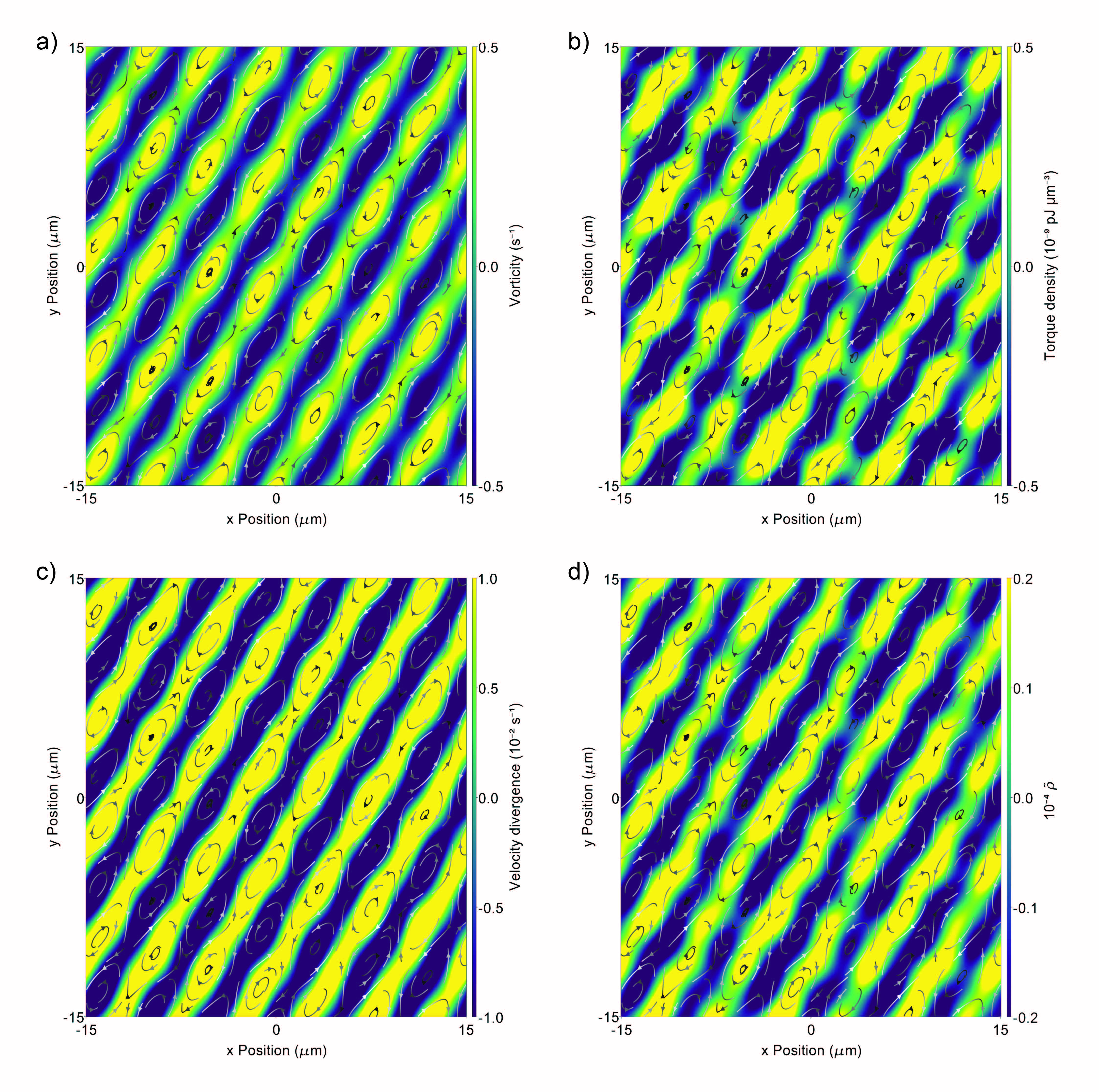}
		\caption{Simulation snapshots of the vorticity $\partial_x v_y - \partial_y v_x$, viscoelastic torque density $\sigma^\text{p}_{xy} - \sigma^\text{p}_{yx} $, velocity divergence $\partial_k v_k$, and density variation $\widetilde{\rho} = (\rho - \rho_\text{s}) / \rho_\text{s}$.  These plots correspond to the same simulation at $t = 0.16\ \text{s}$ for $A = 20\ \text{Pa}$, illustrating that similar spatial patterns have formed for each field.}
		\label{DifferentFields}
	\end{center}
\end{figure}

\clearpage
\begin{figure}[h!]
	\begin{center}
		\includegraphics[width=\textwidth]{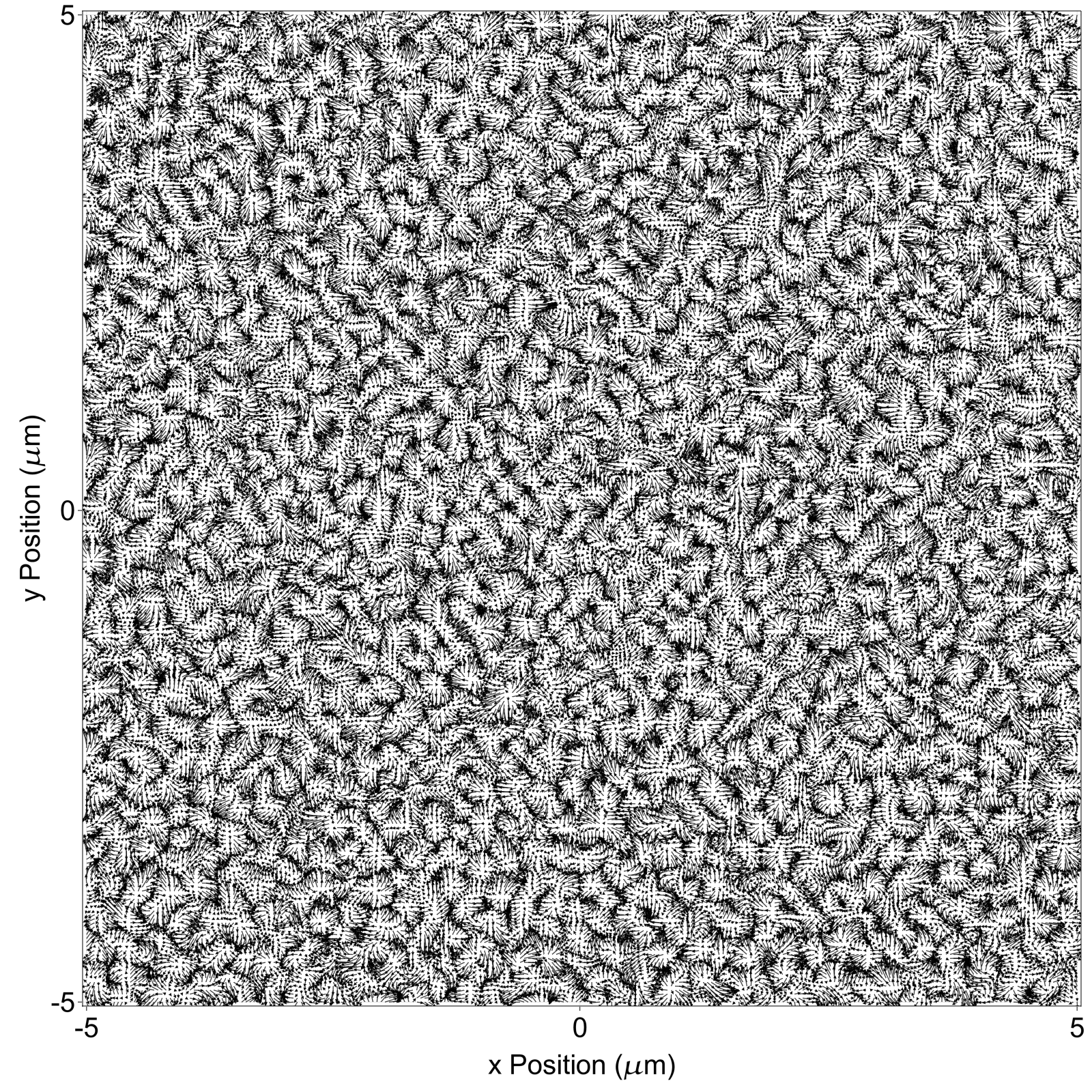}
		\caption{An example vector field generated using Equations \ref{Fxrand} and \ref{Fyrand} is shown.  Here, $N_x = N_y = 250$, $N_f = 300$, and $G = 40$.  For the results reported in the paper, we take $G = 40$ when $N_x = N_y = 250$ and $G = 120$ when $N_x = N_y = 750$.  }
		\label{RandomField}
	\end{center}
\end{figure}

\begin{figure}[h!]
	\begin{center}
		\includegraphics[width=8 cm]{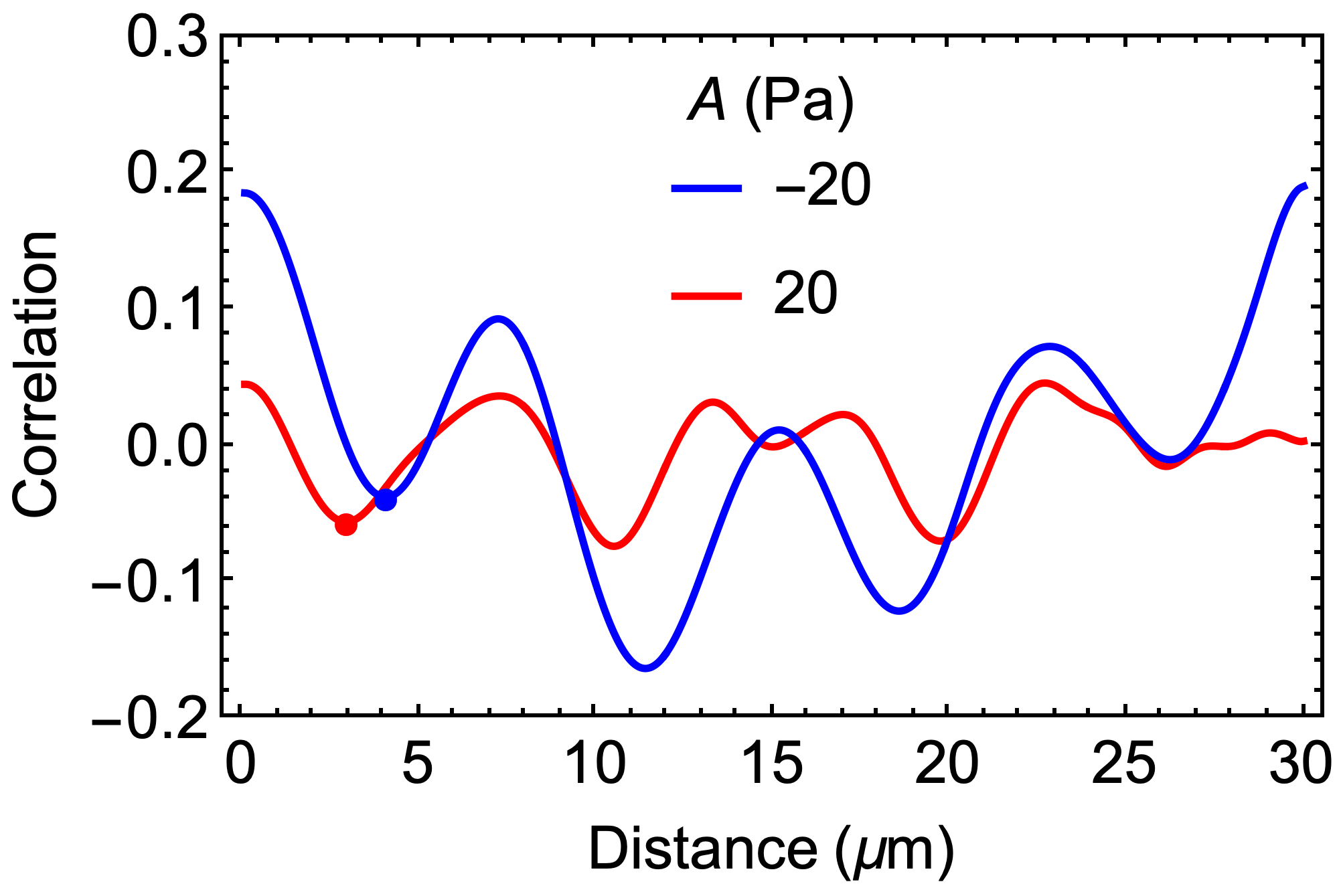}
		\caption{For the simulation snapshots shown Figure 4c,d of the main text, we plot here the measured spatial correlation function of the vorticity field $\Omega(\mathbf{r})$.  From these correlation functions, we take the position of the first minimum (shown as solid circles) a measure of half the spatial wavelength.}
		\label{CorrPlot20}
	\end{center}
\end{figure}

\section{Supplementary Methods}

\subsection{Derivation of odd viscoelastic instability threshold}\label{derivThresh}

Here we derive the linear stability conditions for excitations in an odd viscoelastic fluid. 

The dynamical equations governing the system are
\begin{align}
	\partial_t \rho  &= - \partial_j (\rho v_j) \label{eqmasscons} \\
	p &= c_\text{s}^2 \rho \label{eqp} \\
	\rho D_t v_i &= - \partial_i p + \eta_\text{s} \partial_j \left(\partial_j v_i + \partial_i v_j \right) + \partial_j \sigma_{ij}^\text{p} \label{eqnsa} \\ 
	\mathcal{D}_t\sigma_{ij}^\text{p} &= C_{ijkl}\partial_k v_l - \eta_\text{p}^{-1}C_{ijkl}\sigma_{kl}^\text{p} + D_\text{p} \partial_{kk} \sigma_{ij}^\text{p}. \label{eqsigmapdiff}
\end{align}
See the main text for an explanation of the symbols in these equations.  We have set the extra force density $\mathbf{f}$ to zero here.

In principle the viscous response of the viscoelastic phase may also require a tensorial description $\eta_{\text{p},ijkl}$, causing the second term in Equation \ref{eqsigmapdiff} to depend on a tensor formed from the elasticity and viscosity tensors \cite{banerjee2021active}.  This generalization would not change the form of the dispersion relation derived below, but would require reinterpreting certain coefficients.  We leave this extension to future work.  We also note that one can straightforwardly consider the incompressible case of these dynamics by substituting $p$ for $\rho$ using Equation \ref{eqp} in Equations \ref{eqmasscons} and \ref{eqnsa} and taking the limit $c_\text{s} \rightarrow \infty$.   

We first linearize the above equations around the uniform state $v^{u}_i = 0, \ \rho^{u} = \rho_\text{s}, \ \sigma^{u}_{ij} = 0$, such that $v_i = v_i' \ \rho = \rho_\text{s} + \rho'$, and $\sigma_{ij} = \sigma'_{ij}$, where the primed variables are assumed small.  The material and corotational derivatives reduce to partial derivatives to first order in the small variables.  In what follows we drop the superscript p on the viscoelastic stress tensor, and we also drop the primes, with the understanding that all remaining variables are small.  We eliminate the pressure from Equation \ref{eqnsa} by substituting from Equation \ref{eqp}.  The linearized set of equations is then 
\begin{align}
	\partial_t \rho   &= - \rho_\text{s} \partial_i v_i \label{rhoPartial} \\
	\rho_\text{s} \partial_t v_i &= -c_\text{s}^2 \partial_i \rho + \eta_\text{s}\left(\partial_{jj} v_i + \partial_j \partial_i v_j \right) + \partial_j \sigma_{ij} \label{vPartial} \\
	\partial_t \sigma_{ij} &= - \eta_{p}^{-1} C_{ijkl}\sigma_{kl} + C_{ijkl} \partial_k v_l  + D_\text{p} \partial_{kk} \sigma_{ij}. \label{sigmaPartial}
\end{align}

Next we substitute the general form for the isotropic odd elastic tensor $C_{ijkl}$, which is shown in Ref.\ \citenum{scheibner2020odd} to be
\begin{equation}
	C_{ijkl} = B \delta_{ij}\delta_{kl} + \mu\left(\delta_{il}\delta_{jk} + \delta_{ik}\delta_{jl} - \delta_{ij}\delta_{kl}\right) + K^o E_{ijkl} - A \epsilon_{ij}\delta_{kl} \label{eqCdefagain}
\end{equation}
where 
\begin{equation}
	E_{ijkl} = \frac{1}{2}\left(\epsilon_{ik}\delta_{jl} + \epsilon_{il}\delta_{jk} + \epsilon_{jk}\delta_{il} + \epsilon_{jl}\delta_{ik} \right).
\end{equation}
Here $\delta_{ij}$ is the Kronecker delta and $\epsilon_{ij}$ is the Levi-Civita tensor.  Note that the expression for $C_{ijkl}$ would be symmetric in the indices $i$ and $j$ if not for the term proportional to $A$.  With this, Equation \ref{sigmaPartial} can be written as 
\begin{align}
	\partial_t \sigma_{ij} =& - \eta_{p}^{-1} \left((B- \mu) \delta_{ij} \sigma_{kk} + \mu\left(\sigma_{ij} + \sigma_{ji}\right) - A \epsilon_{ij} \sigma_{kk} + \frac{K^o}{2}\left(\epsilon_{ik}\sigma_{kj} + \epsilon_{il}\sigma_{jl} + \epsilon_{jk}\sigma_{ki} + \epsilon_{jl}\sigma_{il} \right)  \right)  \nonumber \\
	&+ (B- \mu) \delta_{ij} \partial_k v_k + \mu\left(\partial_j v_i + \partial_i v_j \right) - A \epsilon_{ij} \partial_k v_k + \frac{K^o}{2}\left( \epsilon_{ik}\partial_k v_j + \partial_j \epsilon_{ik}v_k + \epsilon_{jk}\partial_k v_i + \partial_i \epsilon_{jk}v_k \right) \nonumber \\
	&+ D_\text{p} \partial_{kk} \sigma_{ij}. \label{eqsigmat}
\end{align}

We next change to the following variables:
\begin{align}
	\Theta &\equiv \partial_i v_i \\
	\Omega &\equiv  \epsilon_{ij}\partial_i v_j \\
	\sigma^I &\equiv \sigma_{ii} \\
	\sigma^{II} &\equiv \partial_i \partial_j \sigma_{ij} \\
	\sigma^{III}_L &\equiv \epsilon_{ik}\partial_j\partial_k \sigma_{ij} \\
	\sigma^{III}_R &\equiv \epsilon_{jk}\partial_i\partial_k \sigma_{ij}. 
\end{align}
Our goal is now to express Equations \ref{rhoPartial}, \ref{vPartial}, and \ref{sigmaPartial} in terms of these new variables.  We divide this process into a few steps, as follows:  
\begin{enumerate}
	
	\item Take the time derivative of Equation \ref{vPartial}. The result of this step is 
	\begin{equation}
		\rho_\text{s} \partial_t^2 v_i = -c_\text{s}^2 \partial_i \partial_t \rho + \eta_\text{s}\partial_t \left( \partial_{jj} v_i + \partial_j \partial_i v_j \right) + \partial_j \partial_t \sigma_{ij}. \label{eqVdouble}
	\end{equation}
	The expression for $\partial_t \sigma_{ij}$ can now be substituted from Equation \ref{eqsigmat}.   
	
	\item Contract Equation \ref{eqVdouble} with $\partial_i$
	This step produces an equation for the time evolution of $\Theta$.  After some algebra, we find
	\begin{align}
		\rho_\text{s} \partial_t^2 \Theta =& -c_\text{s}^2 \partial_t \nabla^2 \rho + 2 \eta_\text{s} \partial_t \nabla^2 \Theta \nonumber \\
		& - \eta_\text{p}^{-1} \left((B-\mu) \nabla^2 \sigma^I + 2 \mu \sigma^{II} - K^o (\sigma^{III}_R + \sigma^{III}_L) \right) \nonumber \\
		&+ (B+\mu) \nabla^2 \Theta + K^o \nabla^2 \Omega + D_\text{p} \nabla^2 \sigma^{II}.
	\end{align}
	
	\item Contract Equation \ref{eqVdouble} with $\epsilon_{ki}\partial_k$.  This step produces an equation for the time evolution of $\Omega$.  We find
	\begin{align}
		\rho_\text{s} \partial_t^2 \Omega =& \eta_\text{s} \partial_t \nabla^2 \Omega \nonumber \\
		&+\eta_\text{p}^{-1} \left(\mu(\sigma_L^{III} + \sigma_R^{III}) - A \nabla^2 \sigma^I + K^o(2 \sigma^{II} - \nabla^2 \sigma^I) \right) \nonumber \\
		&+ \mu \nabla^2 \Omega + A \nabla^2 \Theta - K^o \nabla^2 \Theta - D_\text{p} \nabla^2 \sigma^{III}_L.
	\end{align}
	
	\item Contract Equation \ref{eqsigmat} with $\delta_{ij}$. This step produces an equation for the time evolution of $\sigma^I$.  We find
	\begin{align}
		\partial_t \sigma^I = - \eta_\text{p}^{-1}(B+\mu) \sigma^I + (B + \mu) \Theta + D_\text{p} \nabla^2 \sigma^I.
	\end{align}
	
	\item Contract Equation \ref{eqsigmat} with $\partial_i \partial_j$. This step produces an equation for the time evolution of $\sigma^{II}$.  We find
	\begin{align}
		\partial_t \sigma^{II} =& - \eta_\text{p}^{-1}\left((B-\mu) \nabla^2 \sigma^I + 2 \mu \sigma^{II} - K^o (\sigma_L^{III} + \sigma_R^{III}) \right) \nonumber \\
		&+ (B+\mu) \nabla^2 \Theta + K^o \nabla^2 \Omega + D_\text{p} \nabla^2 \sigma^{II}.
	\end{align}
	
	\item Contract Equation \ref{eqsigmat} with $\epsilon_{ik}\partial_k\partial_j$.  This step produces an equation for the time evolution of $\sigma^{III}_L$.  We find
	\begin{align}
		\partial_t\sigma_L^{III} =& - \eta_\text{p}^{-1}\left( \mu (\sigma_L^{III} + \sigma_R^{III}) - A \nabla^2 \sigma^I + K^o (2 \sigma^{II} - \nabla^2 \sigma^I) \right) \nonumber \\
		&- \mu \nabla^2 \Omega - A \nabla^2 \Theta + K^o \nabla^2 \Theta + D_\text{p} \nabla^2 \sigma_L^{III}.
	\end{align}
	
	\item Contract Equation \ref{eqsigmat} with $\epsilon_{jk}\partial_i\partial_k$.
	This step produces an equation for the time evolution of $\sigma^{III}_R$.  We find
	\begin{align}
		\partial_t\sigma_R^{III} =& - \eta_\text{p}^{-1}\left( \mu (\sigma_R^{III} + \sigma_L^{III}) + A \nabla^2 \sigma^I + K^o (2 \sigma^{II} - \nabla^2 \sigma^I) \right) \nonumber \\
		&- \mu \nabla^2 \Omega + A \nabla^2 \Theta + K^o \nabla^2 \Theta + D_\text{p} \nabla^2 \sigma_R^{III}.
	\end{align}
	
\end{enumerate}

We now have a collection of 7 equations, including $\partial_ t \rho =  - \rho_\text{s} \Theta$ from Equation \ref{rhoPartial}, in 7 variables.  Next, we consider plane wave perturbations corresponding to the ansatz
\begin{equation}
	\Theta(\mathbf{r}, t) \sim \Theta(\mathbf{k}, \omega) e^{\nu t} e^{ i \mathbf{k} \cdot \mathbf{r}}
\end{equation}
and convert the differential equations into algebraic equations in the Fourier coefficients.  After this, we collect everything into the following matrix equation:
\begin{equation}
	\begin{pmatrix}
		\nu & \rho_\text{s} & 0 & 0 & 0 & 0 & 0 \\
		c_\text{s}^2 k^2 \nu & - k^2H_+ - 2k^2 \eta_\text{s} \nu - \rho_\text{s} \nu^2 & -k^2 K^o & \frac{k^2H_-}{\eta_\text{p}} & -k^2D_\text{p} -\frac{2\mu}{\eta_\text{p}} & \frac{K^o}{\eta_\text{p}} &  \frac{K^o}{\eta_\text{p}} \\
		0 & -k^2 G_- & -\rho_\text{s} \nu^2 - k^2 (\mu + \eta_\text{s} \nu) & \frac{k^2G_+}{\eta_\text{p}} & \frac{2 K^o}{\eta_\text{p}} & D_\text{p}k^2 + \frac{\mu}{\eta_\text{p}} & \frac{\mu}{\eta_\text{p}}\\
		0 & H_+ & 0 & -\frac{B +F_1}{\eta_\text{p}} & 0  & 0  & 0\\
		0 & -k^2H_+ & -k^2 K^o & \frac{k^2H_-}{\eta_\text{p}} & -\frac{F_2}{\eta_\text{p}} & \frac{K^o}{\eta_\text{p}} & \frac{K^o}{\eta_\text{p}}\\
		0 & k^2 G_- & k^2 \mu & - \frac{k^2G_+}{\eta_\text{p}} & - \frac{2 K^o}{\eta_\text{p}} & -\frac{F_1}{\eta_\text{p}} & - \frac{\mu}{\eta_\text{p}} \\
		0 & -k^2G_+ & k^2 \mu & \frac{k^2G_-}{\eta_\text{p}} & - \frac{2 K^o}{\eta_\text{p}} & -\frac{\mu}{\eta_\text{p}} & -\frac{F_1 }{\eta_\text{p}}\\
	\end{pmatrix}
	\begin{pmatrix}
		\rho\\
		\Theta\\
		\Omega\\
		\sigma^I\\
		\sigma^{II}\\
		\sigma^{III}_L\\
		\sigma^{III}_R\\
	\end{pmatrix} = 0, \label{eqmatrix}
\end{equation}
where $F_1 \equiv D_\text{p}k^2\eta_\text{p} + \mu + \eta_\text{p}\nu$, $F_2 \equiv D_\text{p}k^2\eta_\text{p} + 2\mu + \eta_\text{p}\nu$, $G_\pm \equiv A \pm K^o$, and $H_{\pm} \equiv B \pm \mu$.

Finally, the dispersion relation $\nu(k)$ is obtained as the solution of $\text{det}(\mathcal{M}) = 0$, where $\mathcal{M}$ is the matrix in Equation \ref{eqmatrix}.  This equation has $9$ solutions which, without any further assumptions, are highly complicated.  We do not write them here, but we note that they reduce to the results derived in Ref.\ \citenum{banerjee2020actin} for the special case $\eta_\text{s} =0, \ \eta_\text{p} \rightarrow \infty, \ D_\text{p} = 0$, and $A = 0$. 

\subsection{Simulation methods}\label{SimMethodsAppendix}

\subsubsection{Numerical algorithm}
To numerically solve the dynamical equations of the odd Jeffrey fluid, we rely on the hybrid lattice Boltzmann (HLB) method using the $d2Q9$ lattice \cite{carenza2019lattice}.  This technique uses a combination of the lattice Boltzmann method to update the velocity and density fields $\mathbf{v}(\mathbf{r})$ and $\rho(\mathbf{r})$, and a finite difference integration scheme to update the polymer orientation vector field $\mathbf{P}(\mathbf{r})$ and the viscoelastic stress tensor field  $\boldsymbol{\sigma}^\text{p}(\mathbf{r})$.  Periodic boundary conditions are used for all fields.

\subsubsection{Initial perturbation}
To numerically study pattern formation we apply a small initial perturbation to the fluid to excite the instability.  This is done through the external body force field $\mathbf{f}_\text{ext}$ which is included as a contribution to the force density $\mathbf{f}$ in Equation 3 of the main text.  Specifically, we apply a force of the form $\mathbf{f}_\text{ext}(\mathbf{r}, t) = T(t) \mathbf{F}^f(\mathbf{r})$, where $T(t)$ is a smooth bump function formed from sigmoidal curves that sets the magnitude of the force field, and $\mathbf{F}^f(\mathbf{r})$ is a normalized vector field.  For the vector field, we use 
\begin{align}
	F_x^f(x,y) =& \frac{1}{H}\sum_{i = 1}^{N_f}\bigg( a^\text{cx}_i \cos(b^\text{cx}_i l_x x + c^\text{cx}_i l_y y) + a^\text{sx}_i \sin(b^\text{sx}_i l_x x + c^\text{sx}_i l_y y)\bigg) \label{Fxrand} \\
	F_y^f(x,y) =& \frac{1}{H}\sum_{i = 1}^{N_f} \bigg(a^\text{cy}_i \cos(b^\text{cy}_i l_x x + c^\text{cy}_i l_y y) + a^\text{sy}_i \sin(b^\text{sy}_i l_x x + c^\text{sy}_i l_y y)\bigg) \label{Fyrand}
\end{align}
where $l_x = 2 \pi/N_x$ and $l_y = 2 \pi/N_y$.  The random numbers $a^\text{cx}_i, \ a^\text{cy}_i, \ a^\text{sx}_i$, and $a^\text{sy}_i$ are drawn from a normal distribution with mean $0$ and variance $1$.  The random numbers $b^\text{cx}_i, \ c^\text{cx}_i, \ b^\text{sx}_i, \ c^\text{sx}_i, \
b^\text{cy}_i, \ c^\text{cy}_i, \ b^\text{sy}_i$, and $ c^\text{sy}_i$ are drawn from a discrete uniform distribution over the domain $[-G, G]$, with the integer $G$ chosen as $120$ for the grid size $N_x = N_y = 750$ and $40$ for the grid size $N_x = N_y = 250$.  We take $N_f = 300$ and then normalize $\mathbf{F}^f(\mathbf{r})$ by choosing $H$ so that the largest vector in the field has unit norm.  The magnitude $T^f(t)$ is kept at $10^{-8}$ (in lattice units) for the first $150$ timesteps of the simulation, with a sigmoidal width of $30$ timesteps.  

The rationale behind this choice of perturbation is that it has the following desirable properties:
\begin{itemize}
	\item It is approximately isotropic, not preferring any direction in the grid.
	\item When $G$ is large it allows for a wide range of spatial frequencies to be excited, increasing the chances that the fastest growing mode of the instability will be excited.
	\item It obeys the periodic boundary conditions, avoiding a large gradient at the boundary due to a discontinuity. 
\end{itemize}

An example of a vector field generated using this method is visualized in Figure \ref{RandomField}.

\subsubsection{Parameterization}

Roughly, the system we have in mind is a micron-scale aqueous viscoelastic solution with elastic moduli on the order of a few $\text{Pa}$, corresponding to the cytoplasm. Experimentally verified parameter values, such as the viscosity of water, were used wherever possible.  Several parameters were instead treated as free rather than constrained by experiments.  The values reported in the tables below are the defaults, so when a given parameter is varied the remaining parameters are set to these values.  We note that, following standard practice with LB simulations, the density of water is set to several orders of magnitude larger than its actual value \cite{tjhung2012spontaneous, cates2004simulating, wolff2012cytoplasmic, henrich2010ordering}.  This allows increasing the time step (thereby speeding up simulations) while still ensuring that the system has a small Reynolds number.  We refer the reader to Ref.\ \citenum{floyd2023simulating} for a full description of all parameters listed below.

\begin{table}[H]
	\begin{center}
		\begin{tabular}{  l | l | l}
			\hline 
			\textbf{Parameter} & \textbf{Symbol} &  \textbf{Value}  \\ \hline 
			Lattice spacing & $\Delta x$  & $4 \times 10^{-8}$ m  \\
			Timestep & $\Delta t$  & $ 8 \times 10^{-6}$ s  \\
			Number of steps & $N_\text{steps}$  & $25,000$, \ $200,000$   \\
			Collision operator time & $\tau$ & 1.25 \\
			Solvent dynamic viscosity & $\eta_\text{s}$  & 0.001 Pa s \\
			Solvent density & $\rho_\text{s}$  & $2 \times 10^7$ kg/m$^3$   \\
			Lattice size & $N_x = N_y$  & $250$, $750$  \\
			\hline
		\end{tabular}
		\caption{Default parameters used in simulation related to the LB algorithm and system domain are shown.  These parameters were used to generate the data on the pattern forming instability.    }\label{LatticeParamsInst}
	\end{center}
\end{table}

\begin{table}[H]
	\begin{center}
		\begin{tabular}{  l | l | l}
			\hline 
			\textbf{Parameter} & \textbf{Symbol} &  \textbf{Value}  \\ \hline
			Polymeric viscosity  & $\eta_\text{p}$ &  0.1 Pa s  \\
			Stress diffusion constant  & $D_\text{p}$ &  $10^{-13}$ m$^2$/s  \\
			Isotropic stiffness tensor element & $B$ & 5.0 Pa \\
			Isotropic stiffness tensor element & $A$ & 5.0 Pa \\
			Isotropic stiffness tensor element & $K^o$ & 5.0 Pa \\
			Isotropic stiffness tensor element & $\mu$ & 2.0 Pa \\
			\hline
		\end{tabular}
		\caption{Default parameters used in simulation related to the viscoelastic stress are shown.  These parameters were used to generate the data on the pattern forming instability. }\label{VEParamsInst}
	\end{center}
\end{table}

\subsubsection{Measuring the wavelength}

To systematically estimate the length scale of the pattern as conditions are varied, we analyze the spatial correlation $C_\Omega(d)$ of the vorticity field $\Omega(\mathbf{r})$.  We first normalize $\Omega$ so that its maximum value over the grid is $1$.  We then estimate $C_\Omega(d)$ as
\begin{equation}
	C_\Omega(d)  = \frac{\sum_{i, j = 1}^{N_x, N_y} \delta(|i - j| - d) \left(\Omega_{ii}\Omega_{ij} +  \Omega_{ii} \Omega_{ji}\right)}{2 \sum_{i,j= 1}^{N_x, N_y} \delta(|i-j| - d)}
\end{equation}
where $\delta(d)$ is the Kronecker delta function, and $\Omega_{ij}$ is value of $\Omega$ at the $i,j$ lattice point.  This formula estimates $C_\Omega(d)$ by evaluating it for reference points along the main diagonal of the grid.  The argument $d = r / \Delta x$  is in lattice units but can be converted to physical units using the simulation length scale $\Delta x$.

For the periodic vortex arrays that make up the typical patterns observed in simulation, the correlation function $C_\Omega(d)$ is also roughly periodic (Figure \ref{CorrPlot20}).  To estimate the wavelength of the array, we pick the value of $d$ where $C_\Omega(d)$ attains its first minimum.  This value of $d$ is then interpreted as half of the pattern's wavelength.   

\color{\editcolor}

\section{A microscopic derivation of the viscoelastic dynamical equations}
Here we consider a tractable microscopic system, a ``non-reciprocal elastic dumbbell'' model, and coarse-grain it following standard procedures to show how new ``odd'' terms emerge alongside those which appear in the usual upper-convected Maxwell model.  This derivation does not reproduce the exact dynamical equations of the odd Jeffreys fluid considered in this paper, which requires a more detailed microscopic model and is left to future work.  However, it does indicate how some new terms which are found in the odd Jeffreys dynamics arise from non-reciprocal interactions at the microscopic level.  

Our derivation primarily follows Refs.\ \citenum{phan2013understanding} and \citenum{larson2013constitutive}, and we consider a 2D system.  The standard elastic dumbbell model was introduced by Kuhn \cite{kuhn1934gestalt} and describes a solution of polymers whose endpoints behave as if connected by a harmonic spring.  The interactions between the polymer and the solvent are localized at these endpoints, which are at $\mathbf{r}_+, \ \mathbf{r}_-$ and are governed by overdamped Langevin dynamics.  The separation vector $\mathbf{R} \equiv \mathbf{r}_+ - \mathbf{r}_-$ and center-of-mass position $\mathbf{R}^c = \frac{1}{2}(\mathbf{r}_+ + \mathbf{r}_-)$ obey, to second order in the solvent velocity $\mathbf{u}^{c}$ evaluated at $\mathbf{R}^{c}$,
\begin{equation}
	\partial_t R_i^{c} = u_i^{c} + \frac{1}{8}R_j R_k \partial_j \partial_k u_i^{c} + \zeta^{-1} F^{b, c}_i(t),
\end{equation}
\begin{equation}
	\partial_t R_i = \left(\partial_j u^c_i\right) R_j + 2\zeta^{-1} F_i^{\text{int}} + \zeta^{-1}F^{b}_i(t). \label{eqRi}
\end{equation}
Here, $\zeta$ is a scalar friction coefficient obeying the fluctuation-dissipation relation with the Brownian forces $\mathbf{F}^{b,c}$ and $\mathbf{F}^{b}$ which act, respectively, on $\mathbf{R}^c$ and $\mathbf{R}$:
\begin{equation}
	\left \langle F_i^{b,c}(t) \right \rangle = 0, \ \ \ \left \langle F_i^{b,c}(t+t')F_j^{b,c}(t) \right \rangle = k_B T \zeta \delta(t') \delta_{ij},
\end{equation}
\begin{equation}
	\left \langle F_i^{b}(t) \right \rangle = 0, \ \ \ \left \langle F_i^{b}(t+t')F_j^{b}(t) \right \rangle = 4k_B T \zeta \delta(t') \delta_{ij}.
\end{equation}
In the standard elastic dumbbell model, the interaction force is \begin{equation}
	F_i^{\text{int}} = - k_\parallel R_i,
\end{equation}
i.e., the polymer behaves like a harmonic spring with zero rest length.  We add to this interaction a transverse force depending on the separation $\mathbf{R}$:
\begin{equation}
	F_i^{\text{int}} = - k_\parallel R_i - k_\perp \epsilon_{ij} R_j = -M_{ij} R_j,
\end{equation}
where 
\begin{equation}
	M_{ij} = k_\parallel \delta_{ij} + k_\perp \epsilon_{ij}.
\end{equation}

We seek the evolution of the quantity $\left \langle R_i R_j \right \rangle$, which we will eventually relate to the viscoelastic stress tensor $\sigma_{ij}^\text{p}$.  Using Equation \ref{eqRi}, we have
\begin{equation}
	\partial_t(R_i R_j) = \left(\partial_k u_i^c\right)R_k R_j + R_i R_k\left(\partial_k u^c_j \right) + 2 \zeta^{-1}\left(F^\text{int}_iR_j + F^\text{int}_jR_i\right) + \zeta^{-1}\left(F_i^b(t)R_j + F_j^b(t)R_i\right). \label{eqRR}
\end{equation}
In expectation, the Brownian force terms can be simplified using a separation of timescales between $R_i$ and $F_i^b(t)$ \cite{phan2013understanding}:
\begin{equation}
	\left \langle R_i F_j^b \right \rangle \approx \left \langle F_i^b R_j \right \rangle \approx 2k_BT\delta_{ij} .
\end{equation}
The interaction terms can be expressed as 
\begin{equation}
	F^\text{int}_iR_j + F^\text{int}_jR_i = -M_{ik}R_k R_j - R_iM_{jk}R_k = -\widetilde{C}_{ijkl} R_k R_l
\end{equation}
where
\begin{align}
	\widetilde{C}_{ijkl} =& M_{ik}\delta_{jl} + M_{jk}\delta_{il} \\
	=&k_\parallel\left(\delta_{ik}\delta_{jl} + \delta_{jk}\delta_{il} \right) + \frac{k_\perp}{2}\left(\epsilon_{ik}\delta_{jl} + \epsilon_{il}\delta_{jk} + \epsilon_{jk}\delta_{il} + \epsilon_{jl}\delta_{ik}\right).
\end{align}
This tensor is of the same form as Equation \ref{eqCdefagain} above, when $B = \mu$ and $A = 0$.  We note that the symmetrized combinations $(\epsilon_{ik}\delta_{jl} + \epsilon_{il}\delta_{jk})/2$ and $(\epsilon_{jk}\delta_{il} + \epsilon_{jl}\delta_{ik})/2$ appear because the contraction with $R_k R_l$ should be symmetric under the interchange of indices $k$ and $l$.  The expectation of Equation \ref{eqRR} can now be written as
\begin{equation}
	\partial_t \left \langle R_i R_j \right \rangle  = \left(\partial_k u_i^c\right) \left \langle R_k R_j \right \rangle + \left \langle R_i R_k \right \rangle \left(\partial_k u^c_j \right) - 2 \zeta^{-1}\widetilde{C}_{ijkl} \left \langle R_k R_l\right \rangle + 4 k_B T \zeta^{-1} \delta_{ij}.\label{eqRRF}
\end{equation}
The polymer-contributed stress can be written as $S^\text{p}_{ij} = (1/2)\nu \widetilde{C}_{ijkl}\left \langle R_k R_l \right \rangle$ where $\nu$ is the number density of polymers.  Contracting both sides of Equation \ref{eqRRF} with $\nu\widetilde{C}_{ijkl}$ and using the upper convected derivative $\mathcal{D}_t^\text{uc}X_{ij} = \partial_t X_{ij} - \left(\partial_k u_i^c\right) X_{kj}  - X_{ik} \left(\partial_k u^c_j \right)$, we have
\begin{equation}
	\nu\widetilde{C}_{ijkl} \mathcal{D}_t^\text{uc} \left \langle R_k R_l \right \rangle = - 4\zeta^{-1}\widetilde{C}_{ijkl}S^\text{p}_{kl} + 4 k_B T \nu\zeta^{-1} \widetilde{C}_{ijkl} \delta_{kl}.
\end{equation}
The operations of $\mathcal{D}^\text{uc}_t$ and contraction with $\widetilde{C}_{ijkl}$ do not commute when $k_\perp \neq 0$. However, since we are interested in the linear stability regime, for simplicity, we neglect the commutator of these operations, which is second order in the perturbations around the homogeneous state.  It is standard to redefine the polymer stress to absorb the pressure-like term:
$\sigma_{ij}^\text{p} = S_{ij}^\text{p} - k_B T \nu \delta_{ij}$.  With this, we can write
\begin{equation}
	\mathcal{D}^\text{uc}_tS_{ij}^\text{p} = \mathcal{D}^\text{uc}_t\sigma_{ij}^\text{p} + k_BT\nu \mathcal{D}^\text{uc}_t \delta_{ij} = -2\zeta^{-1}\widetilde{C}_{ijkl}\sigma_{kl}^\text{p}.
\end{equation}
One can show that $\mathcal{D}^\text{uc}_t \delta_{ij}  = -\left(\partial_i u_j^c + \partial_j u_i^c\right) = - 2 \Psi_{ij}$, so that
\begin{equation}
	\mathcal{D}^\text{uc}_t\sigma_{ij}^\text{p} = 2 k_BT\nu \Psi_{ij} - 2 \zeta^{-1}\widetilde{C}_{ijkl}\sigma_{kl}^\text{p}. \label{equca}
\end{equation}
If $k_\perp = 0$, then Equation \ref{equca} reduces to the standard upper convected Maxwell model:
\begin{equation}
	\mathcal{D}_t^\text{uc}\sigma_{ij}^\text{p} = 2k_BT\nu\Psi_{ij} - 4k_\parallel\zeta^{-1} \sigma_{ij}^\text{p}.
\end{equation}
Defining the relaxation time $\lambda = \zeta / 4k_\parallel $ and thermal energy density $G = \nu k_B T$, this can be expressed in the familiar form 
\begin{equation}
	\lambda \mathcal{D}_t^\text{uc}\sigma_{ij}^\text{p} = 2 \widetilde{\eta}_\text{p}\Psi_{ij} + \sigma_{ij}^\text{p}
\end{equation}
where $\widetilde{\eta}_\text{p} = G \lambda$.  If the polymeric elasticity is due to the Kuhn stiffness of a Gaussian chain, we can write
\begin{equation}
	k_\parallel = a k_B T,
\end{equation}
where the proportionality $a$ (having units of inverse length squared) is related to the Kuhn length and the number of Kuhn segments.  By assuming a similar relationship between stiffness $\widetilde{C}_{ijkl}$ and thermal energy for the case when $k_\perp \neq 0$, we generalize the coefficient of $\Psi_{ij}$ in Equation \ref{equca} so that it reads
\begin{equation}
	\mathcal{D}_t^\text{uc}\sigma_{ij}^\text{p} = \frac{2 \nu}{a}\widetilde{C}_{ijkl}\Psi_{ij} -2\zeta^{-1}\widetilde{C}_{ijkl}\sigma_{kl}^\text{p}.
\end{equation}  
Defining $C_{ijkl} = \frac{2 \nu}{a} \widetilde{C}_{ijkl}$ and $\eta_\text{p} = \zeta \nu /a$, we can write
\begin{equation}
	\mathcal{D}_t^\text{uc}\sigma_{ij}^\text{p} = C_{ijkl}\Psi_{ij} -\eta_\text{p}^{-1}C_{ijkl}\sigma_{kl}^\text{p}.
\end{equation}
This last equation can be compared to Equation \ref{eqsigmapdiff}, and $\sigma_{ij}^\text{p}$ can be added to the solvent phase stresses to give the final Navier-Stokes equation for the Oldroyd-B fluid.  An Oldroyd-B fluid, which is the same as a Jeffrey fluid but with an upper-convected rather than co-rotational derivative in the evolution for the viscoelastic stress tensor, is a model for a viscoelastic Maxwell material immersed in a viscous solvent.  The two components are assumed to flow together, and thus one velocity field suffices to describe the motion of the combined system.  The two components however make separate additive contributions to the total stress tensor, which enters the momentum-balance encoded in the Navier-Stokes equation.  This physical approximation to the two-component system of polymer and solvent is quite standard in rheology, and models like the Jeffrey fluid have successfully reproduced numerous experimental observations \cite{phan2013understanding, bird1987dynamics, larson2013constitutive}.  We note that in certain situations this model suffers from physically inconsistent singularities owing to the assumed infinite extensibility of the constituent polymers.  For our linear instability calculations this issue is not important.  

The derivation presented here does not exactly reproduce Equation \ref{eqsigmapdiff}, missing some terms in the elasticity tensor and using a different materially objective derivative, but it illustrates how the odd coefficient $K^o$ in the elasticity tensor, which drives instabilities, arises from microscopic forces proportional to $k_\perp$.  Future work could focus on a similar derivation of viscoelastic dynamics from spinning colloids, rather than from a polymeric system with transverse forces as considered here.  

\color{black}

\bibliographystyle{unsrt}

\end{document}